\documentclass[11pt, a4paper]{article}
\pdfoutput = 1
\usepackage{jheppub}
\usepackage{graphicx, slashed, bbm, makecell, feynmp-auto}

\allowdisplaybreaks

\begin{document}

\title{\Large Vector-Like Quarks and Leptons, SU(5) $\otimes$ SU(5) Grand Unification, and Proton Decay}

\author{Chang-Hun Lee and Rabindra N.~Mohapatra}
\affiliation{Maryland Center for Fundamental Physics and Department of Physics, University of Maryland, College Park, Maryland 20742, USA}

\begin{abstract}
{SU(5) $\otimes$ SU(5) provides a minimal grand unification scheme for fermions and gauge forces if there are vector-like quarks and leptons in nature. We explore the gauge coupling unification in a non-supersymmetric model of this type, and study its implications for proton decay. The properties of vector-like quarks and intermediate scales that emerge from coupling unification play a central role in suppressing proton decay. We find that in this model, the familiar decay mode $p \to e^+ \pi^0$ may have a partial lifetime within the reach of currently planned experiments.}
\end{abstract}

\maketitle

\section{Introduction}	\label{s.intro}
Grand unification of forces and matter~\cite{GUT} is a very attractive idea, and has been a popular venue for exploring new physics beyond the Standard Model (SM) for the past four decades. A key requirement of these theories is that two electroweak couplings and the strong coupling of the SM (denoted by their fine-structure constants $\alpha_Y$, $\alpha_w$, and $\alpha_s$), which are known very precisely at the weak scale $\mu = M_Z$, become equal when extrapolated according to the effective field theories at energies below the grand unification scale. Since there is generally a large gap between $M_Z$ and the typical scales of grand unified theories (GUT) in different realizations of unification, a lot of unknown physics could exist in between: in fact, as such is suggested by the fact that the SM gauge couplings with only SM particles do not unify. This leaves open the possibility that many other phenomena, not explained in the SM such as the observed Higgs mass or neutrino masses, could be playing a role in gauge coupling unification. A famous and widely discussed example is the naturalness problem of Higgs mass, one solution to which is supersymmetry (SUSY) at the weak scale. It is well known that adding SUSY brings in new bosonic and fermionic states near the TeV scale, which miraculously leads to coupling unification without any further gauge symmetry or new particles in between~\cite{drw}. The simplest grand unification group in this case is SUSY SU(5). However, minimal versions of this model have problems not only with predictions for proton decay~\cite{mp}, but also in accommodating fermion masses. Nevertheless, coupling unification as a probe of physics beyond the SM has sustained its excitement over the years, although lack of experimental evidence for SUSY in the Large Hadron Collider (LHC) has somewhat dampened the interest in the SUSY-GUT possibility and has led to a revival of interest in coupling unification schemes that do not involve SUSY.

A widely discussed alternative to SUSY SU(5) grand unification is the one based on the SO(10) group~\cite{so10} where coupling unification can be achieved without SUSY in a way different from the minimal SU(5) case and therefore has a much richer history than the former. It was originally inspired by two observations: (a) the basic \textbf{16}-dimensional spinor fits all the fermions of one generation together with the right-handed (RH) neutrino and provides minimal unification of fermions per generation; (b) it naturally predicts that neutrinos have nonzero masses. Activity in this field therefore ramped up after the discovery of neutrino oscillations confirming neutrino masses in 1998 and its possible explanation via the seesaw mechanism~\cite{type1}. In this case, grand unification can proceed via single or multiple intermediate states unlike SU(5), and detailed analysis of unification of gauge couplings in this model has been the subject of many papers for different scenarios~\cite{so10u}. Another advantage of non-SUSY SO(10) over SU(5) is the possibility of low intermediate scales as well as new observable physics in currently available particle physics facilities. An additional advantage of these models is that they included as a subgroup the left-right symmetric gauge symmetry~\cite{LR} according to which parity symmetry becomes an exact symmetry of weak interactions at short distance scales.

In this paper, we consider another route to unification of forces and matter without SUSY based on the gauge group SU(5) $\otimes$ SU(5) $\otimes~Z_2$~\cite{su5su5}. This is the minimal unification group if there are vector-like quarks and leptons at scales above TeV. This also provides an alternative GUT embedding of left-right symmetry like SO(10).
An interesting property of these models is that masses of light quarks and charged leptons arise via the seesaw mechanism with the vector-like quarks and leptons providing the heavy mass scale, as we discuss below~\cite{quarkseesaw}.
The vector-like fermion is an essential part of this model, and it is one of the major new physics search goals of the LHC. In fact, if vector-like fermions are discovered in colliders, it will strongly point towards a grand unification based on the SU(5) $\otimes$ SU(5) group. The reason is that while other groups such as SU(5) or SO(10) can accommodate vector-like fermions once additional multiplets beyond the minimal setup are included, SU(5) $\otimes$ SU(5) $\otimes~Z_2$ unification is the minimal unification group that predicts their existence and no other fermions. There are also other motivations to consider product groups $G$ $\otimes$ $G$ based on the possibility of a hidden sector for dark matter and its unification~\cite{hidden}. We do not go this route in this paper. While such models have been discussed in literature~\cite{su5su5}, no realistic non-SUSY grand unified model has been presented.
 
The main point of this paper is to address the issue of gauge coupling unification without SUSY in the minimal SU(5) $\otimes$ SU(5) model.  We present one scenario where coupling unification is successfully achieved at the one-loop level. Other scenarios may be possible, but we focus only on one case which appears consistent with all low energy observations and discuss its predictions for proton lifetime. We find it interesting that while full coupling unification does occur at around $10^{16}$ GeV, one of the SU(5) groups does unify at around $10^{11}$ GeV. This does lead to proton decay via the exchange of gauge bosons with masses around $10^{11}$ GeV. Even though such a low unification scale has the potential to lead to a very short proton lifetime, the original operator involves heavy vector-like fermions and the actual proton decay arises via the mixing of the heavy and light fermions. This together with matrix element suppression from QCD calculations leads to a prediction for proton lifetime at or above the current experimental lower limits from Super-Kamiokande~\cite{SuperK} experiment. The most interesting of these is the prediction for $p \to e^+ \pi^0$ which is near $6 \cdot 10^{34}$ yrs., and should be testable in the next round of experiments like Hyper-Kamiokande~\cite{hyperK}. The simplest model also predicts a Dirac neutrino, in which case there are decay modes $p \to \nu_e^c \pi^+$ and $p \to \nu_\mu^c K^+$ and the proton lifetime through these channels are also in the range of $3 \cdot 10^{34}$ yrs. We must caution that these estimates have uncertainties that we do not address here, such as those from two-loop effects or threshold corrections due to Higgs mass distributions around intermediate scales. These uncertainties are unlikely to shift the proton lifetime by more than an order of magnitude, which means that any improvement of the proton lifetime bound at upcoming experiments like Hyper-Kamiokande will severely constrain these models.
 
 
This paper is organized as follows: in Sec.~\ref{sec:model}, we present the basic elements of the model; Sec.~\ref{sec:gc} is devoted to the discussion of gauge coupling unification; Sec.~\ref{sec:proton} presents the implications of our model for proton lifetime; in Sec.~\ref{sec:comment}, some comments and the summary of the results are provided. The appendix~\ref{appendix} is devoted to a study of the scalar potential and the scalar mass spectrum.

\section{Outline of the model} \label{sec:model}
The fermions in this model are assigned to $(\overline{\textbf{5}}, \textbf{1}) \oplus (\textbf{10}, \textbf{1}) \oplus (\textbf{1}, \textbf{5}) \oplus (\textbf{1}, \overline{\textbf{10}})$ representations of SU(5)$_A$ $\otimes$ SU(5)$_B$ as follows (all fields are assumed to have left-handed (LH) chirality as in usual discussions of GUT):
\begin{align}
	\psi &= \left( \begin{array}{c} D_1^c \\ D_2^c \\ D_3^c \\ e^- \\ \nu \end{array} \right) \sim (\overline{\textbf{5}}, \textbf{1}), \\
	\chi &= \left( \begin{array}{ccccc}
		0 & U_3^c & -U_2^c & u_1 & d_1 \\
		-U_3^c & 0 & U_1^c & u_2 & d_2 \\
		U_2^c & -U_1^c & 0 & u_3 & d_3 \\
		-u_1 & -u_2 & -u_3 & 0 & E^+ \\
		-d_1 & -d_2 & -d_3 & -E^+ & 0 \end{array} \right) \sim (\textbf{10}, \textbf{1}),
\end{align}
and
\begin{align}
	\psi^c &= \left( \begin{array}{c} D_1 \\ D_2 \\ D_3 \\ e^+ \\ \nu^c \end{array} \right) \sim (\textbf{1}, \textbf{5}), \label{eq:fermion5} \\
	\chi^c &= \left( \begin{array}{ccccc}
		0 & U_3 & -U_2 & u_1^c & d_1^c \\
		-U_3 & 0 & U_1 & u_2^c & d_2^c \\
		U_2 & -U_1 & 0 & u_3^c & d_3^c \\
		-u_1^c & -u_2^c & -u_3^c & 0 & E^- \\
		-d_1^c & -d_2^c & -d_3^c & -E^- & 0 \end{array} \right) \sim (\textbf{1}, \overline{\textbf{10}}).
\end{align}
There are three copies of these multiplets corresponding to three generations. Note that the new heavy fermions of the model $(U, D, E)$ are singlets of the LH and RH subgroups SU(2)$_{A/B}$ of SU(5)$_{A/B}$ groups, and are therefore vector-like. Note that the RH neutrino appears naturally in this model by group theory requirement.

\subsection{Scalar fields}
The GUT gauge group breaks down to the SM gauge groups after the Higgs multiplets acquire vacuum expectation values (VEV). We choose the following scalar multiplets for this purpose, and their SU(5)$_A$ $\otimes$ SU(5)$_B$ representations are given by
\begin{gather}
	H_A \sim (\textbf{5}, \textbf{1}), \qquad
	H_B \sim (\textbf{1}, \textbf{5}), \qquad
	S \sim (\overline{\textbf{10}}, \textbf{10}), \qquad
	S' \sim (\overline{\textbf{10}}, \textbf{10}), \nonumber \\
	\Phi \sim (\textbf{5}, \overline{\textbf{5}}), \qquad
	\Sigma_A \sim (\textbf{24}, \textbf{1}), \qquad
	\Sigma_B \sim (\textbf{1}, \textbf{24}).
\end{gather}
We present the Higgs potential and its minimization in Appendix \ref{appendix}. The VEV's needed for our purpose are
\begin{align}
	\langle H_k \rangle &= (0, 0, 0, 0, v_{H_k})^\mathsf{T}, \\
	\langle {S_{ij}}^{\alpha \beta} \rangle &= \left\{ \begin{array}{ll} (-1)^{p + q} v_S & \text{if}~\{i, j; \alpha, \beta\} = \{p(4, 5); q(4, 5)\}, \\ 0 & \text{otherwise}, \end{array} \right. \\
	\langle {S'_{ij}}^{\alpha \beta} \rangle &= 0, \\
	\langle \Phi \rangle &= v_\Phi \text{diag}(1, 1, 1, 0, 0), \\
	\langle \Sigma_k \rangle &= v_{\Sigma_k} \text{diag}(2, 2, 2, -3, -3)
\end{align}
where $k = A, B$ and $p, q$ are some permutations of given indices. Here, $S'$ is introduced for providing extra scalar multiplets needed for coupling unification at the GUT scale. The intermediate scales of the model above the weak scale $M_Z \sim |v_{H_A}|$ are $M_R \sim |v_{H_B}|$, $M_S \sim |v_S|$, $M_\Phi \sim |v_\Phi|$, $M_\Sigma \sim v_{\Sigma_B}$, and $M_\text{GUT} \sim v_{\Sigma_A}$. The spontaneous symmetry breaking chain in our model is given in (\ref{eq:ssbchain}), with multiplets that are responsible for the corresponding breaking also given next to the downarrows.
\begin{align}
	\text{SU}(5)_A &\otimes \text{SU}(5)_B \otimes Z_2 \nonumber \\
		&\downarrow \ M_\text{GUT} \sim \langle \Sigma_A \rangle \nonumber \\
	\text{SU}(3)_A &\otimes \text{SU}(2)_A \otimes \text{U}(1)_A \otimes \text{SU}(5)_B \nonumber \\
		&\downarrow \ M_\Sigma \sim \langle \Sigma_B \rangle \nonumber \\
	\text{SU}(3)_A &\otimes \text{SU}(2)_A \otimes \text{U}(1)_A \otimes \text{SU}(3)_B \otimes \text{SU}(2)_B \otimes \text{U}(1)_B \nonumber \\
		&\downarrow \ M_{\Phi} \sim \langle \Phi \rangle \nonumber \\
	\text{SU}(3)_s &\otimes \text{SU}(2)_A \otimes \text{U}(1)_ A \otimes \text{SU}(2)_B \otimes \text{U}(1)_B \label{eq:ssbchain} \\
		&\downarrow \ M_S \sim \langle S \rangle \nonumber \\
	\text{SU}(3)_s &\otimes \text{SU}(2)_A \otimes \text{SU}(2)_B \otimes \text{U}(1)_{B - L} \nonumber \\
		&\downarrow \ M_R \sim \langle H_B \rangle \nonumber \\
	\text{SU}(3)_s &\otimes \text{SU}(2)_A \otimes \text{U}(1)_Y \nonumber \\
		&\downarrow \ M_Z \sim \langle H_A \rangle \nonumber \\
	\text{SU}(3)_s &\otimes \text{U}(1)_\text{em} \nonumber
\end{align}
The SU(5) scalar multiplets are decomposed into SU(3) $\otimes$ SU(2) submultiplets as follows:
\begin{align}
	H &\to H^T \oplus H^D, \\
	S &\to {S_{TT}}^{TT} \oplus {S_{TT}}^{TD} \oplus {S_{TT}}^{DD} \oplus {S_{TD}}^{TT} \oplus {S_{TD}}^{TD} \oplus {S_{TD}}^{DD} \nonumber \\
	&\qquad \qquad \oplus {S_{DD}}^{TT} \oplus {S_{DD}}^{TD} \oplus {S_{DD}}^{DD} \qquad ({S_{TT}}^{TT} = {S_{\bar{3}}}^3, \ {S_{DD}}^{DD} = {S_1}^1), \\
	\Phi &\to \Phi^{TT} \oplus \Phi^{TD} \oplus \Phi^{DT} \oplus \Phi^{DD} \qquad (\Phi^{TT} = \Phi^T_8 \oplus \Phi^T_1, \ \Phi^{DD} = \Phi^D_3 \oplus \Phi^D_1), \\
	\Sigma &\to \Sigma^{TT} \oplus \Sigma^{TD} \oplus \Sigma^{DT} \oplus \Sigma^{DD} \nonumber \\
	&\qquad \qquad (\Sigma^{TT} = \Sigma^T_8 \oplus \Sigma^T_1, \ \Sigma^{DD} = \Sigma^D_3 \oplus \Sigma^D_1, \ \Sigma_1 \equiv \Sigma^T_1 = -\Sigma^D_1).
\end{align}
Here, ``$T$'' and ``$D$'' denote SU(3) and SU(2) indices, respectively. In case of $\Phi$ and $\Sigma$, the subscript ``8'' is for an SU(3) octet, ``3'' for an SU(2) triplet, and ``1'' for a singlet. Similarly, in case of $S$, ``3'' is for the SU(3) triplet, and ``1'' for a singlet. The gauge transformation properties of scalar submultiplets are summarized in Table \ref{tab:mass}. The scalar fields that acquire nonzero VEV's are $\phi_{H_A^D}$ in $H_A^D$, $\phi_{H_B^D}$ in $H_B^D$, ${S_1}^1$, $\Phi^T_1$, $\Sigma_{A1}$, and $\Sigma_{B1}$, and the mass spectrum of scalar submultiplets are given in Appendix \ref{appendix}.

\subsection{Yukawa Lagrangian}
In this model, the Yukawa coupling terms responsible for giving masses to the fermions are
\begin{align}
	\mathcal{L}_Y &= -h_{H_A} (H_A^\dagger)_i (\chi^{ij})^\mathsf{T} \mathsf{C} \psi_j
	+ \frac{1}{8} h'_{H_A} \epsilon_{ijk \ell m} {H_A}^i (\chi^{jk})^\mathsf{T} \mathsf{C} \chi^{\ell m} \nonumber \\
	&\qquad - h_{H_B} {H_B}^\alpha ({\chi^c}_{\alpha \beta})^\mathsf{T} \mathsf{C} {\psi^c}^\beta
	+ \frac{1}{8} h'_{H_B} \epsilon^{\alpha \beta \gamma \delta \lambda} (H_B^\dagger)_\alpha ({\chi^c}_{\beta \gamma})^\mathsf{T} \mathsf{C} {\chi^c}_{\delta \lambda} \nonumber \\
	&\qquad + h_\Phi {\Phi^i}_\alpha ({\psi^c}^\alpha)^\mathsf{T} \mathsf{C} \psi_i
	+ \frac{1}{4} h_S {S_{ij}}^{\alpha \beta} ({\chi^c}_{\alpha \beta})^\mathsf{T} \mathsf{C} \chi^{ij}
	+ \frac{1}{4} h_{S'} {S'_{ij}}^{\alpha \beta} ({\chi^c}_{\alpha \beta})^\mathsf{T} \mathsf{C} \chi^{ij} \nonumber \\
	&\qquad + \text{H.c.}.
\end{align}
Here, Roman indices are for SU(5)$_A$ and Greek indices for SU(5)$_B$, and $\mathsf{C}  = i \gamma^2 \gamma^0$ is the charge conjugation operator in the Dirac-Pauli representation of $\gamma^\mu$'s. We have suppressed the fermion generation indices.
Now we introduce a $Z_2$ symmetry under which the fermionic and bosonic fields transform as
\begin{align}
	\psi \leftrightarrow \psi^{c *}, \quad
	\chi \leftrightarrow \chi^{c *}, \quad
	H_A \leftrightarrow H_B, \quad
	\Sigma_A \leftrightarrow \Sigma_B, \quad
	\Phi \leftrightarrow \Phi^\dagger, \quad
	S \leftrightarrow S^\dagger.
\end{align}
Invariance of the Lagrangian under this symmetry relates the Yukawa couplings of the model as follows:
\begin{align}
	\mathcal{L}_Y &= -h_H (H_A^\dagger)_i (\chi^{ij})^\mathsf{T} \mathsf{C} \psi_j
	+ \frac{1}{8} h'_H \epsilon_{ijk \ell m} {H_A}^i (\chi^{jk})^\mathsf{T} \mathsf{C} \chi^{\ell m} \nonumber \\
	&\qquad - h_H^* {H_B}^\alpha ({\chi^c}_{\alpha \beta})^\mathsf{T} \mathsf{C} {\psi^c}^\beta
	+ \frac{1}{8} h_H'^* \epsilon^{\alpha \beta \gamma \delta \lambda} (H_B^\dagger)_\alpha ({\chi^c}_{\beta \gamma})^\mathsf{T} \mathsf{C} {\chi^c}_{\delta \lambda} \nonumber \\
	&\qquad + h_\Phi {\Phi^i}_\alpha ({\psi^c}^\alpha)^\mathsf{T} \mathsf{C} \psi_i
	+ \frac{1}{4} h_S {S_{ij}}^{\alpha \beta} ({\chi^c}_{\alpha \beta})^\mathsf{T} \mathsf{C} \chi^{ij}
	+ \frac{1}{4} h_{S'} {S'_{ij}}^{\alpha \beta} ({\chi^c}_{\alpha \beta})^\mathsf{T} \mathsf{C} \chi^{ij} \nonumber \\
	&\qquad + \text{H.c.}
	\label{eq:YP}
\end{align}
where $h_\Phi, h_S, h_S'$ are Hermitian matrices in the generation space.

\subsection{Fermion masses}
After spontaneous symmetry breaking, the charged fermion mass matrices are given in the seesaw forms and this induces small mixing between light and heavy fermions, which we show here. This small mixing suppresses proton decay, as we will see in Sec.~\ref{sec:proton}. To explain  the direct heavy fermion masses in the seesaw matrix, we note that  unlike the vector-like fermions $D$ and $E$ whose masses directly come from the Yukawa Lagrangian (\ref{eq:YP}), the $U$ quark mass arises from the effective Lagrangian term
\begin{align}
	h_S^{ij} \lambda_U U_i^\mathsf{T} \mathsf{C} U_j^c
\end{align}
generated by the tadpole diagrams
\begin{center}
	\unitlength = 1 mm
\begin{fmffile}{UquarkFD}
	\qquad \qquad
	\parbox{25 mm}{
		\tiny
		\begin{fmfgraph*}(25,18)
			\fmfstraight
			\fmfset{arrow_len}{2 mm}
			\fmfset{wiggly_len}{2 mm}
			
			\fmfleft{i1,xi2,xi3}
			\fmfright{o1,xo2,xo3}
			
			\fmf{fermion,width=0.7}{i1,v1}
			\fmf{fermion,width=0.7}{v1,o1}
			\fmffreeze
			
			\fmf{dashes,width=0.7,label=${S_{23}}^{23}$,label.side=right,label.dist=1.5}{v1,v2}
			\fmf{phantom}{xi2,v3}
			\fmf{dashes,width=0.7,tension=0.7}{v3,v2,v4}
			\fmf{phantom}{v4,xo2}
			
			\fmfv{label=$\frac{1}{4} h_S^{ij}$,label.angle=-90,label.dist=2}{v1}
			\fmfv{label=$\eta_{\Phi S}$,label.angle=-150,label.dist=1.5}{v2}
			
			\fmfv{label=$(\chi_i^c)_{23} = (U_i)_1$,label.dist=1}{i1}
			\fmfv{label=$(\chi_j)^{23} = (U_j^c)_1$,label.dist=1}{o1}
			\fmfv{label=$\langle {(\Phi^\dagger)^2}_2 \rangle = v_\Phi^*$,label.angle=120,label.dist=1.3,decor.shape=circle,decor.size=3}{v3}
			\fmfv{label=$\langle {(\Phi^\dagger)^3}_3 \rangle = v_\Phi^*$,label.angle=60,label.dist=1.3,decor.shape=circle,decor.size=3}{v4}
		\end{fmfgraph*}}
		\qquad \qquad \qquad \qquad \qquad \quad
	\parbox{25 mm}{
		\tiny
		\begin{fmfgraph*}(25,18)
			\fmfstraight
			\fmfset{arrow_len}{2 mm}
			\fmfset{wiggly_len}{2 mm}
			
			\fmfleft{i1,xi2,xi3}
			\fmfright{o1,xo2,xo3}
			
			\fmf{fermion,width=0.7}{i1,v1}
			\fmf{fermion,width=0.7}{v1,o1}
			\fmffreeze
			
			\fmf{dashes,width=0.7,label=${S_{23}}^{23}$,label.side=right,label.dist=1.5}{v1,v2}
			\fmf{phantom}{xi2,v3}
			\fmf{dashes,width=0.7,tension=0.7}{v3,v2,v4}
			\fmf{phantom}{v4,xo2}
			
			\fmfv{label=$\frac{1}{4} h_S^{ij}$,label.angle=-90,label.dist=2}{v1}
			\fmfv{label=$\eta_{S \Phi}$,label.angle=-150,label.dist=1.5}{v2}
			
			\fmfv{label=$(\chi_i^c)_{23} = (U_i)_1$,label.dist=1}{i1}
			\fmfv{label=$(\chi_j)^{23} = (U_j^c)_1$,label.dist=1}{o1}
			\fmfv{label=$\langle {\Phi^1}_1 \rangle = v_\Phi$,label.angle=120,label.dist=1.3,decor.shape=circle,decor.size=3}{v3}
			\fmfv{label=$\langle {(S^\dagger)_{45}}^{45} \rangle = v_S^*$,label.angle=60,label.dist=1.3,decor.shape=circle,decor.size=3}{v4}
		\end{fmfgraph*}} \newline \newline
\end{fmffile}
\end{center}
where $i, j$ are generation indices and we have shown only one specific choice of SU(5)$_A$ $\otimes$ SU(5)$_B$ indices for simplicity. The scalar potential terms responsible for these diagrams can be found in Appendix \ref{appendix}. After spontaneous symmetry breaking, the parameter $\lambda_U$ is given by
\begin{align}
	\lambda_U \sim \frac{\eta_{\Phi S} v_\Phi^{*2}}{4 m_{{S_{TT}}^{TT}}^2}
		+ \frac{\eta_{S \Phi} v_\Phi v_S^*}{4 m_{{S_{TT}}^{TT}}^2}
\end{align}
at the energy scale $\mu \ll m_{{S_{TT}}^{TT}}$. Now the fermion mass terms after symmetry breaking are
\begin{align}
	\mathcal{L}_f &= (\textbf{d}^\mathsf{T} \ \textbf{D}^\mathsf{T}) \mathsf{C}
		\left( \begin{array}{cc} 0 & h_H v_{H_A}^* \\ h_H^\dagger v_{H_B} & h_\Phi v_\Phi \end{array} \right)
		\left( \begin{array}{c} \textbf{d}^c \\ \textbf{D}^c \end{array} \right)
	+ (\textbf{u}^\mathsf{T} \ \textbf{U}^\mathsf{T}) \mathsf{C}
		\left( \begin{array}{cc} 0 & h'_H v_{H_A} \\ h_H'^\dagger v_{H_B}^* & h_S \lambda_U \end{array} \right)
		\left( \begin{array}{c} \textbf{u}^c \\ \textbf{U}^c \end{array} \right) \nonumber \\
	&\qquad + (\textbf{e}^{- \mathsf{T}} \ \textbf{E}^{- \mathsf{T}}) \mathsf{C}
		\left( \begin{array}{cc} 0 & h_H^\mathsf{T} v_{H_A}^* \\ h_H^* v_{H_B} & h_S v_S \end{array} \right)
		\left( \begin{array}{c} \textbf{e}^+ \\ \textbf{E}^+ \end{array} \right)
	+ \text{H.c.}
	\label{eq:fmass}
\end{align}
where
\begin{alignat}{3}
	&\textbf{u} = (u, c, t)^\mathsf{T}, \qquad
	&&\textbf{d} = (d, s, b)^\mathsf{T}, \qquad
	&&\textbf{e} = (e, \mu, \tau)^\mathsf{T}, \nonumber \\
	&\textbf{U} = (U_1, U_2, U_3)^\mathsf{T}, \qquad
	&&\textbf{D} = (D_1, D_2, D_3)^\mathsf{T}, \qquad
	&&\textbf{E} = (E_1, E_2, E_3)^\mathsf{T}.
\end{alignat}
Note that entries in different fermion mass matrices share common Yukawa couplings and VEV's.

The general structure of quark and charged lepton mass matrices is
\begin{align}
	M = \left( \begin{array}{cc} 0 & h v_A \\ h^\dagger v_B & g v_C \end{array} \right)
\end{align}
where $g$ is Hermitian. When $|h^{ij} v_A| \ll |h^{kl} v_B| \ll |g^{mn} v_C|$, we can approximately write the light and heavy fermion mass matrices $M_f$, $M_F$ as
\begin{align}
	M_f \approx h g^{-1} h^\dagger U_f \frac{v_A v_B}{v_C}, \qquad
	M_F \approx g U_F v_C
	\label{eq:fFmass}
\end{align}
where $U_f$ and $U_F$ are some unitary matrices.
Note that the light fermion mass matrix $M_f$ is given in the seesaw form: the large diagonal term $g v_C$ and the relatively small off-diagonal terms $h v_A$, $h^\dagger v_B$ generate small light fermion masses.

Since the mass eigenstate of a light fermion is mainly composed of the light fermion in the generation basis when $|h^{ij} v_A| \ll |h^{kl} v_B| \ll |g^{mn} v_C|$, we can write $\textbf{f}_m \approx \textbf{f}_g + V_f \textbf{F}_g$ $(\textbf{f} = \textbf{u}, \textbf{d}, \textbf{e}$, $\textbf{F} = \textbf{U}, \textbf{D}, \textbf{E})$ where $m$ and $g$ denote mass and generation bases, respectively. Here, $V_f$ is the $3 \times 3$ mixing matrix between $\textbf{f}_m$ and $\textbf{F}_g$, and it is written as
\begin{align}
	V_f \approx -h g^{-1} \frac{v_A}{v_C}.
	\label{eq:mixing}
\end{align}
This mixing is indeed small when $|h^{ij} v_A| \ll |g^{mn} v_C|$, which is a crucial factor that allows this model to be viable in spite of the constraints from proton lifetime, as we see later.

In fact, these mass and mixing matrices are only roughly correct for real masses, since the actual Yukawa coupling matrices typically have hierarchy in their entries, i.e.~the condition $|h^{ij} v_A| \ll |h^{kl} v_B| \ll |g^{mn} v_C|$ is not satisfied for all the entries of the matrices.

For neutrino masses, we also introduce the effective Lagrangian term
\begin{align}
	m_\nu^{ij} \nu_i^\mathsf{T} \mathsf{C} \nu^c_j
\end{align}
induced by the Feynman diagrams
\begin{center}
	\unitlength = 1 mm
\begin{fmffile}{neutrinoFD}
	\quad \qquad
	\parbox{25 mm}{
		\tiny
		\begin{fmfgraph*}(25,18)
			\fmfstraight
			\fmfset{arrow_len}{2 mm}
			\fmfset{wiggly_len}{2 mm}
			
			\fmfleft{i1,xi2,xi3}
			\fmfright{o1,xo2,xo3}
			
			\fmf{fermion,width=0.7}{i1,v1}
			\fmf{fermion,width=0.7}{v1,o1}
			\fmffreeze
			
			\fmf{dashes,width=0.7,label=${\Phi^5}_5$,label.side=right,label.dist=1.5}{v1,v2}
			\fmf{phantom}{xi2,v3}
			\fmf{dashes,width=0.7,tension=0.7}{v3,v2,v4}
			\fmf{phantom}{v4,xo2}
			
			\fmfv{label=$h_\Phi^{ij}$,label.angle=-90,label.dist=2}{v1}
			\fmfv{label=$\eta_{H \Phi}$,label.angle=-150,label.dist=1.5}{v2}
			
			\fmfv{label=$(\psi_i)_5 = \nu_i$,label.dist=1}{i1}
			\fmfv{label=$(\psi_j^c)^5 = \nu_j^c$,label.dist=1}{o1}
			\fmfv{label=$\langle (H_A)^5 \rangle = v_{H_A}$,label.angle=120,label.dist=1.3,decor.shape=circle,decor.size=3}{v3}
			\fmfv{label=$\langle (H_B^\dagger)_5 \rangle = v_{H_B}^*$,label.angle=60,label.dist=1.3,decor.shape=circle,decor.size=3}{v4}
		\end{fmfgraph*}}
		\qquad \qquad \qquad \qquad \qquad \quad
	\parbox{25 mm}{
		\tiny
		\begin{fmfgraph*}(25,18)
			\fmfstraight
			\fmfset{arrow_len}{2 mm}
			\fmfset{wiggly_len}{2 mm}
			
			\fmfleft{i1,xi2,xi3,xi4,xi5,xi6,xi7}
			\fmfright{o1,xo2,xo3,xo4,xo5,xo6,xo7}
			
			\fmf{fermion,width=0.7}{i1,v1}
			\fmf{fermion,width=0.7}{v1,o1}
			\fmffreeze
			
			\fmf{dashes,width=0.7,label=${\Phi^5}_5$,label.side=right,label.dist=1.5}{v1,v2}
			\fmf{phantom}{xi3,v3}
			\fmf{dashes,width=0.7,tension=0.7}{v3,v2,v4}
			\fmf{phantom}{v4,xo3}
			\fmf{phantom}{xi4,v5,xo4}
			\fmf{dashes,width=0.7,tension=0.7}{v2,v5}
						
			\fmfv{label=$h_\Phi^{ij}$,label.angle=-90,label.dist=2}{v1}
			\fmfv{label=$\lambda_{H \Sigma \Phi}^*$,label.angle=-150,label.dist=1.5}{v2}
			
			\fmfv{label=$(\psi_i)_5 = \nu_i$,label.dist=1}{i1}
			\fmfv{label=$(\psi_j^c)^5 = \nu_j^c$,label.dist=1}{o1}
			\fmfv{label=$\langle (H_A)^5 \rangle = v_{H_A}$,label.angle=120,label.dist=1.3,decor.shape=circle,decor.size=3}{v3}
			\fmfv{label=$\langle (H_B^\dagger)_5 \rangle = v_{H_B}^*$,label.angle=60,label.dist=1.3,decor.shape=circle,decor.size=3}{v4}
			\fmfv{label=$\langle {(\Sigma_A)^5}_5 \rangle = -3 v_{\Sigma_A}$,label.angle=90,label.dist=1.3,decor.shape=circle,decor.size=3}{v5}
		\end{fmfgraph*}} \newline \newline
\end{fmffile}
\end{center}
The components of neutrino mass matrix after spontaneous symmetry breaking are thus given by
\begin{align}
	m_\nu^{ij} \sim \frac{h_\Phi^{ij} \eta_{H \Phi} v_{H_A} v_{H_B}^*}{m_{{\Phi^5}_5}^2}
		- \frac{3 h_\Phi^{ij} \lambda_{H \Sigma \Phi}^* v_{H_A} v_{H_B}^* v_{\Sigma_A}}{m_{{\Phi^5}_5}^2}
\end{align}
at $\mu \ll m_{{\Phi^5}_5}$. Note that neutrinos are Dirac-type in this model.

\section{Gauge coupling unification} \label{sec:gc}
Let us now turn to the gauge coupling unification. For simplicity, we assume that any fields with masses larger than an energy scale $\mu$ completely decouple from the theory at and below $\mu$. In addition, we only consider one-loop $\beta$-functions, in which case the fine-structure constants at $\mu$ satisfy
\begin{align}
	\alpha^{-1} (\mu) = \alpha^{-1} (\mu_0) + \frac{b}{2\pi} \ln{(\mu / \mu_0)}.
\end{align}
Furthermore, we ignore any threshold corrections with the exception of $H_B^T$ mass. In order to suppress the proton decay mediated by $H_B^T$, it turns out that the mass of $H_B^T$ should be larger than the threshold $M_\Sigma$, the unification scale of SU(5)$_B$.

In general, the threshold corrections are implemented in a matching condition as follows: when a gauge group $G$ breaks into several gauge groups $G_i$'s, gauge couplings at $\mu$ satisfy the renormalization group equation~\cite{hall} 
\begin{align}
	\alpha_i^{-1} (\mu) &= \alpha_G^{-1} (\mu) - \frac{\lambda_i (\mu)}{12\pi}
\end{align}
where
\begin{align}
	\lambda_i (\mu) = (C_G - C_i) - 21 \text{tr} \big[ t_{iV}^2 \ln{(M_V/\mu)} \big] + \text{tr} \big[ t_{iS}^2 P_\text{GB} \ln{(M_S/\mu)} \big] + 8 \text{tr} \big[ t_{iF}^2 \ln{(M_F/\mu)} \big].
\end{align}
Here, $V$, $F$, and $S$ denote heavy vector gauge bosons, fermions, and scalar fields contributions, respectively, to the internal loops of the field strength renormalization diagrams of external gauge bosons of $G_i$. $C_G$ and $C_i$ are quadratic Casimir invariants of $G$ and $G_i$, respectively, and $t_{iV}$, $t_{iS}$, and $t_{iF}$ are generators of $G_i$ corresponding to heavy internal fields. $P_{GB}$ is the projection operator that projects out Goldstone bosons.

In order to write the matching conditions at various symmetry breaking scales, we need to know the relationships among the group generators. For this purpose, we write $\lambda^a / 2$ $(a = 1, \cdots, 8)$ and $\sigma^b / 2$ $(b = 1, 2, 3)$ as SU(3) and SU(2) generators, respectively, and $V_{A/B} / 2$ as the generators of U(1)$_{A/B}$ groups. Then, we have
\begin{align}
	\frac{\lambda_s^a}{2} &= \frac{\lambda_A^a}{2} + \frac{\lambda_B^a}{2}, \\
	\frac{B - L}{2} &= \frac{V_A}{2} + \frac{V_B}{2}, \\
	\frac{Y}{2} &= \sqrt{\frac{3}{13}} \frac{\sigma_B^3}{2} + \sqrt{\frac{5}{13}} \frac{B - L}{2}
\end{align}
where $\lambda_s^a/2$ are the generators of the strong force gauge group SU(3)$_s$. Note that most generators are canonically normalized, e.g.~$\text{tr}(\lambda^a \lambda^b) = 2 \delta^{ab}$. The exceptions are $\lambda_s^a / 2$ and $(B - L) / 2$, the generators of diagonal subgroups  SU(3)$_{A + B}$ = SU(3)$_s$ and U(1)$_{A + B}$ = U(1)$_{B - L}$.\footnote{In principle, these generators can be written as linear combinations such as $\lambda_s^a / 2 = k (\lambda_{A}^a / 2 + \lambda_{B}^a / 2)$ for an arbitrary constant $k$. However, $k = 1$ is the right choice if we require the coupling constant $g_s$ associated with $\lambda_s^a$ be the strong coupling constant of the SM. This follows from the facts that (i) the SU(3) generators of the SM as well as $\lambda_{A/B}^a / 2$ are canonically normalized, and that (ii) the SU(3)$_s$ gauge boson field strength renormalization diagram with a chiral fermion loop have the same value as that of SU(3)$_{A/B}$ when $\lambda_s^a = \lambda_{A}^a + \lambda_{B}^a$. A similar argument applies to $B - L$.} Also note that $Y = \sqrt{3/13} Y'$ and $B - L = \sqrt{3/5} (B - L)'$ where $Y'/2$ is the weak hypercharge of the SM and $(B - L)'$ is the baryon number minus the lepton number.

\begin{figure}[h]
	\centering
	\includegraphics[width = 0.9 \textwidth]{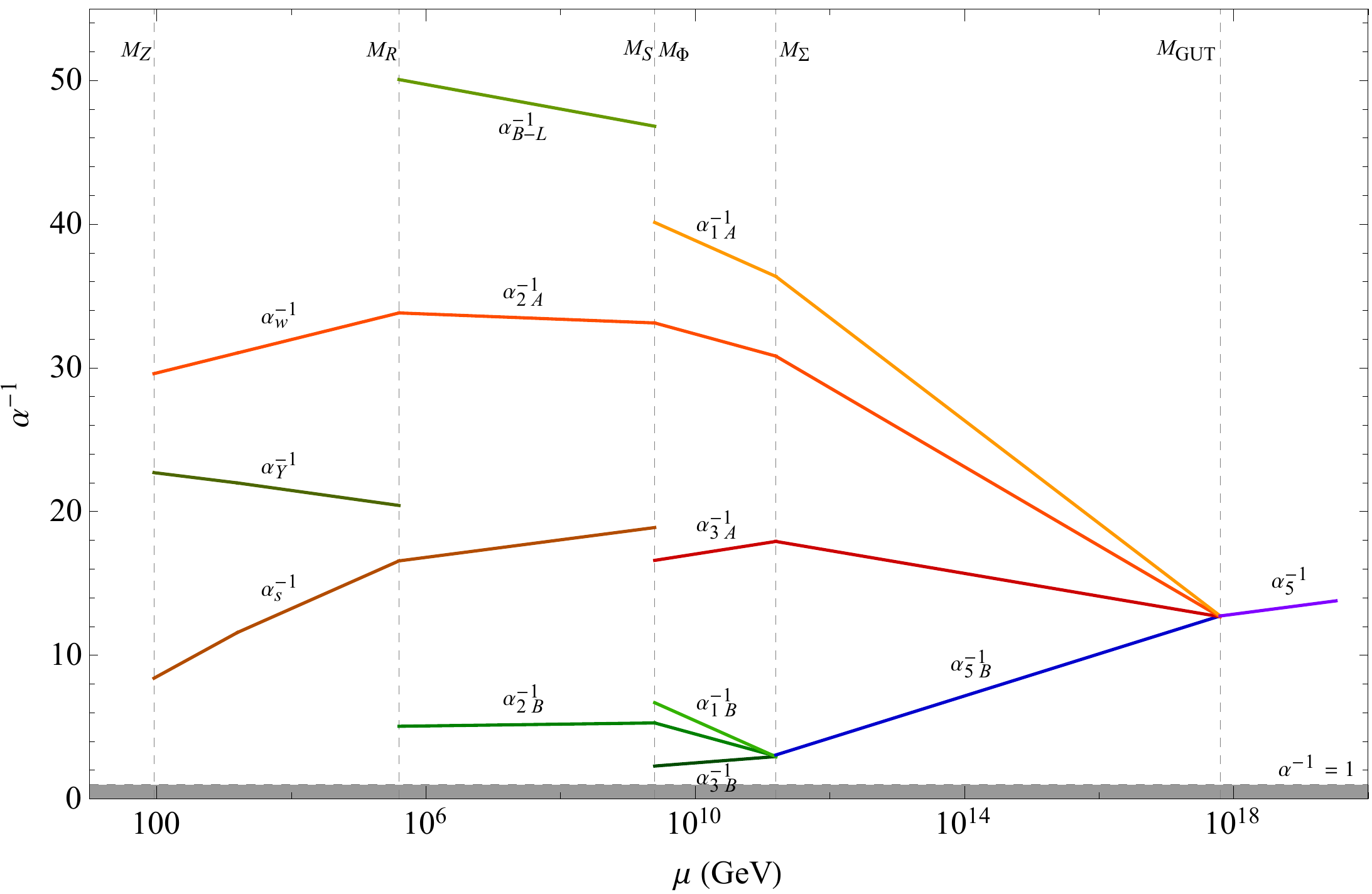}
	\caption{Gauge coupling unification.}
	\label{fig:GUT}
\end{figure}

The matching conditions at each boundary of energy scales are then given by~\cite{matching}
\begin{align}
	\alpha_Y^{-1} (M_R) &= \frac{3}{13} \alpha_{2B}^{-1} (M_R) + \frac{5}{13} \alpha_{B - L}^{-1} (M_R), \\
	\alpha_{B - L}^{-1} (M_S) &= \alpha_{1A}^{-1} (M_S) + \alpha_{1B}^{-1} (M_S), \\
	\alpha_s^{-1} (M_\Phi) &= \alpha_{3A}^{-1} (M_\Phi) + \alpha_{3B}^{-1} (M_\Phi), \\
	\alpha_{5BT}^{-1} (M_\Sigma) &= \alpha_{3B}^{-1} (M_\Sigma) + \frac{1}{12 \pi} \bigg[ 2 + \frac{1}{2} \ln{(m_{H_B^T}/M_\Sigma)} \bigg] \nonumber \\
		&= \alpha_{2B}^{-1} (M_\Sigma) + \frac{1}{4 \pi} \nonumber \\
		&= \alpha_{1B}^{-1} (M_\Sigma) + \frac{1}{12 \pi} \bigg[ 5 - \frac{1}{15} \ln{(m_{H_B^T}/M_\Sigma)} \bigg], \label{eq:alpha5B} \\
	\alpha_5^{-1} (M_\text{GUT}) &= \alpha_{3A}^{-1} (M_\text{GUT}) = \alpha_{2A}^{-1} (M_\text{GUT}) = \alpha_{2A}^{-1} (M_\text{GUT})
\end{align}
where ``T'' in $\alpha_{5BT}^{-1}$ of (\ref{eq:alpha5B}) represents the inclusion of threshold effects. The plot of gauge coupling unification is given in Fig.~\ref{fig:GUT}. For simplicity, we have assumed $M_S = M_\Phi$ since the more general choice $M_S < M_\Phi$ turned out to be no better at all for any purposes. We have also used numerical matching conditions $\frac{3}{13} \alpha_{2B}^{-1} (M_R) = 0.0571 \alpha_Y^{-1} (M_R)$, $\alpha_{1A}^{-1} (M_S) = 0.857 \alpha_{B - L}^{-1} (M_S)$, and $\alpha_{3A}^{-1} (M_\Phi) = 0.879 \alpha_s^{-1} (M_\Phi)$. Note that we have $\alpha_{5BT}^{-1} (M_\Sigma) = 3.056$, which is the value used in calculations of proton decay rates in Sec.~\ref{sec:proton}.
The masses of heavy fermions and scalar submultiplets that lead to the unification of gauge couplings are given in Table \ref{tab:mass}.
\begin{table}[h]
	\small
	\begin{tabular}{|c||c|c|}
		\hline
		Heavy fermion & Representation & Mass \\ \hline
		$U_1$ & $(\textbf{1}, \textbf{1}, 0; \textbf{3}, \textbf{1}, \frac{4}{3})$ & $M_S$ \\ \hline
		$U_1^c$ & $(\textbf{3}, \textbf{1}, -\frac{4}{3}; \textbf{1}, \textbf{1}, 0)$ & $M_S$ \\ \hline
		$U_2$ & $(\textbf{1}, \textbf{1}, 0; \textbf{3}, \textbf{1}, \frac{4}{3})$ & $M_S$ \\ \hline
		$U_2^c$ & $(\textbf{3}, \textbf{1}, -\frac{4}{3}; \textbf{1}, \textbf{1}, 0)$ & $M_S$ \\ \hline
		$U_3$ & $(\textbf{1}, \textbf{1}, 0; \textbf{3}, \textbf{1}, \frac{4}{3})$ & $M_S$ \\ \hline
		$U_3^c$ & $(\textbf{3}, \textbf{1}, -\frac{4}{3}; \textbf{1}, \textbf{1}, 0)$ & $M_S$ \\ \hline
		$D_1$ & $(\textbf{1}, \textbf{1}, 0; \textbf{3}, \textbf{1}, -\frac{2}{3})$ & $M_D$ \\ \hline
		$D_1^c$ & $(\textbf{3}, \textbf{1}, \frac{2}{3}; \textbf{1}, \textbf{1}, 0)$ & $M_D$ \\ \hline
		$D_2$ & $(\textbf{1}, \textbf{1}, 0; \textbf{3}, \textbf{1}, -\frac{2}{3})$ & $M_D$ \\ \hline
		$D_2^c$ & $(\textbf{3}, \textbf{1}, \frac{2}{3}; \textbf{1}, \textbf{1}, 0)$ & $M_D$ \\ \hline
		$D_3$ & $(\textbf{1}, \textbf{1}, 0; \textbf{3}, \textbf{1}, -\frac{2}{3})$ & $M_\Phi$ \\ \hline
		$D_3^c$ & $(\textbf{3}, \textbf{1}, \frac{2}{3}; \textbf{1}, \textbf{1}, 0)$ & $M_\Phi$ \\ \hline
		$E_1^-$ & $(\textbf{1}, \textbf{1}, 0; \textbf{1}, \textbf{1}, -2)$ & $M_S$ \\ \hline
		$E_1^+$ & $(\textbf{1}, \textbf{1}, 2; \textbf{1}, \textbf{1}, 0)$ & $M_S$ \\ \hline
		$E_2^-$ & $(\textbf{1}, \textbf{1}, 0; \textbf{1}, \textbf{1}, -2)$ & $M_S$ \\ \hline
		$E_2^+$ & $(\textbf{1}, \textbf{1}, 2; \textbf{1}, \textbf{1}, 0)$ & $M_S$ \\ \hline
		$E_3^-$ & $(\textbf{1}, \textbf{1}, 0; \textbf{1}, \textbf{1}, -2)$ & $M_S$ \\ \hline
		$E_3^+$ & $(\textbf{1}, \textbf{1}, 2; \textbf{1}, \textbf{1}, 0)$ & $M_S$ \\ \hline
	\end{tabular}
	\begin{tabular}{|c||c|c|}
		\hline
		Scalar submultiplet & Representation & Mass \\ \hline
		$H_A^T$ & $(\textbf{3}, \textbf{1}, -\frac{2}{3}; \textbf{1}, \textbf{1}, 0)$ & $M_\text{GUT}$ \\ \hline
		$H_A^D$ & $(\textbf{1}, \textbf{2}, 1; \textbf{1}, \textbf{1}, 0)$ & $M_Z$ \\ \hline
		$H_B^T$ & $(\textbf{1}, \textbf{1}, 0; \textbf{3}, \textbf{1}, -\frac{2}{3})$ & $10^{1.8} M_\Sigma$ \\ \hline
		$H_B^D$ & $(\textbf{1}, \textbf{1}, 0; \textbf{1}, \textbf{2}, 1)$ & $M_R$ \\ \hline
		$\Phi^{TT}$ & $(\textbf{3}, \textbf{1}, -\frac{2}{3}; \overline{\textbf{3}}, \textbf{1}, \frac{2}{3})$ & $M_\Phi$ \\ \hline
		$\Phi^{TD}$ & $(\textbf{3}, \textbf{1}, -\frac{2}{3}; \textbf{1}, \textbf{2}, -1)$ & $M_\Sigma$ \\ \hline
		$\Phi^{DT}$ & $(\textbf{1}, \textbf{2}, 1; \overline{\textbf{3}}, \textbf{1}, \frac{2}{3})$ & $M_\text{GUT}$ \\ \hline
		$\Phi^{DD}$ & $(\textbf{1}, \textbf{2}, 1; \textbf{1}, \textbf{2}, -1)$ & $M_\text{GUT}$ \\ \hline
		${S_{TT}}^{TT}$ & $(\overline{\textbf{3}}, \textbf{1}, \frac{4}{3}; \textbf{3}, \textbf{1}, -\frac{4}{3})$ & $M_S$ \\ \hline
		${S_{TT}}^{TD}$ & $(\overline{\textbf{3}}, \textbf{1}, \frac{4}{3}; \textbf{3}, \textbf{2}, \frac{1}{3})$ & $M_\Sigma$ \\ \hline
		${S_{TT}}^{DD}$ & $(\overline{\textbf{3}}, \textbf{1}, \frac{4}{3}; \textbf{1}, \textbf{1}, 2)$ & $M_\Sigma$ \\ \hline
		${S_{TD}}^{TT}$ & $(\overline{\textbf{3}}, \textbf{2}, -\frac{1}{3}; \textbf{3}, \textbf{1}, -\frac{4}{3})$ & $M_\Sigma$ \\ \hline
		${S_{TD}}^{TD}$ & $(\overline{\textbf{3}}, \textbf{2}, -\frac{1}{3}; \textbf{3}, \textbf{2}, \frac{1}{3})$ & $M_R$ \\ \hline
		${S_{TD}}^{DD}$ & $(\overline{\textbf{3}}, \textbf{2}, -\frac{1}{3}; \textbf{1}, \textbf{1}, 2)$ & $M_\Sigma$ \\ \hline
		${S_{DD}}^{TT}$ & $(\textbf{1}, \textbf{1}, -2; \textbf{3}, \textbf{1}, -\frac{4}{3})$ & $M_\Sigma$ \\ \hline
		${S_{DD}}^{TD}$ & $(\textbf{1}, \textbf{1}, -2; \textbf{3}, \textbf{2}, \frac{1}{3})$ & $M_\Sigma$ \\ \hline
		${S_{DD}}^{DD}$ & $(\textbf{1}, \textbf{1}, -2; \textbf{1}, \textbf{1}, 2)$ & $M_S$ \\ \hline
		${S'_{TD}}^{TT}$ & $(\overline{\textbf{3}}, \textbf{2}, -\frac{1}{3}; \textbf{3}, \textbf{1}, -\frac{4}{3})$ & $M_\Sigma$ \\ \hline
		${S'_{TD}}^{TD}$ & $(\overline{\textbf{3}}, \textbf{2}, -\frac{1}{3}; \textbf{3}, \textbf{2}, \frac{1}{3})$ & $M_\Phi$ \\ \hline
		${S'_{TD}}^{DD}$ & $(\overline{\textbf{3}}, \textbf{2}, -\frac{1}{3}; \textbf{1}, \textbf{1}, 2)$ & $M_\Sigma$ \\ \hline
		$\Sigma_A^{TT}$ & $(\textbf{3}, \textbf{1}, 0; \textbf{1}, \textbf{1}, 0)$ & $M_\text{GUT}$ \\ \hline
		$\Sigma_A^{TD}$ & $(\textbf{3}, \textbf{2}, -\frac{5}{3}; \textbf{1}, \textbf{1}, 0)$ & $M_\text{GUT}$ \\ \hline
		$\Sigma_A^{DT}$ & $(\textbf{3}, \textbf{2}, \frac{5}{3}; \textbf{1}, \textbf{1}, 0)$ & $M_\text{GUT}$ \\ \hline
		$\Sigma_A^{DD}$ & $(\textbf{2}, \textbf{2}, 0; \textbf{1}, \textbf{1}, 0)$ & $M_R$ \\ \hline
		$\Sigma_B^{TT}$ & $(\textbf{1}, \textbf{1}, 0; \textbf{3}, \textbf{1}, 0)$ & $M_\Phi$ \\ \hline
		$\Sigma_B^{TD}$ & $(\textbf{1}, \textbf{1}, 0; \textbf{3}, \textbf{2}, -\frac{5}{3})$ & $M_\Sigma$ \\ \hline
		$\Sigma_B^{DT}$ & $(\textbf{1}, \textbf{1}, 0; \textbf{3}, \textbf{2}, \frac{5}{3})$ & $M_\Sigma$ \\ \hline
		$\Sigma_B^{DD}$ & $(\textbf{1}, \textbf{1}, 0; \textbf{2}, \textbf{2}, 0)$ & $M_\Phi$ \\ \hline
	\end{tabular}
	\caption{The properties of heavy fermions and scalar submultiplets. The representation is for the gauge group SU(3)$_A \otimes$ SU(2)$_A \otimes$ U(1)$_A \otimes$ SU(3)$_B \otimes$ SU(2)$_B \otimes$ U(1)$_B$. Note that, for simplicity, we have presented the values of $(B - L)' = Y' - \sigma_w^3$ where $Y' / 2$ is the weak hypercharge of the SM and $\sigma_w^3 / 2$ is one of the SU(2)$_w$ generators. The rest of the submultiplets in $S'$ have masses of $M_\text{GUT}$. Also note that we have presented would-be Goldstone bosons with the masses of gauge bosons that absorb them.}
	\label{tab:mass}
\end{table}
The energy scales and VEV's as well as heavy fermions and scalar submultiplets associated with each energy scale are presented in Table \ref{tab:Escale}.
\begin{table}
	\center
	\begin{tabular}{|c||c|c||c|c|}
		\hline
		Energy scale & VEV & Value (GeV) & Heavy fermion & Scalar submultiplet \\ \hline
		$M_Z$ & $v_{H_A}$ & 91.2 & $\cdot$ & $H_A^D$ \\ \hline
		$M_D$ & $\cdot$ & $10^{3.2}$ & $D_1$, $D_1^c$, $D_2$, $D_2^c$ & $\cdot$ \\ \hline
		$M_R$ & $v_{H_B}$ & $10^{5.6}$ & $\cdot$ & $H_B^D$, ${S_{TD}}^{TD}$, $\Sigma_A^{DD}$ \\ \hline
		$M_S$ & $v_S$ & $10^{9.4}$ &	\makecell{$U_1$, $U_1^c$, $U_2$, $U_2^c$, \\ $U_3$, $U_3^c$, $E_1^-$, $E_1^+$, \\ $E_2^-$, $E_2^+$, $E_3^-$, $E_3^+$} & ${S_{TT}}^{TT}$, ${S_{DD}}^{DD}$ \\ \hline
		$M_\Phi$ & $v_\Phi$ & $10^{9.4}$ & $D_3$, $D_3^c$ & ${S'_{TD}}^{TD}$, $\Phi^{TT}$, $\Sigma_B^{TT}$, $\Sigma_B^{DD}$ \\ \hline
		$M_\Sigma$ & $v_{\Sigma_B}$ & $10^{11.2}$ & $\cdot$ & 	\makecell{$\Phi^{TD}$, ${S_{TT}}^{TD}$, ${S_{TT}}^{DD}$, \\ ${S_{TD}}^{TT}$, ${S_{TD}}^{DD}$, ${S_{DD}}^{TT}$, \\ ${S_{DD}}^{TD}$, ${S'_{TD}}^{TT}$, ${S'_{TD}}^{DD}$, \\ $\Sigma_B^{TD}$, $\Sigma_B^{DT}$} \\ \hline
		$(10^{1.8} M_\Sigma)$ & $\cdot$ & $10^{13}$ & $\cdot$ & $H_B^T$ \\ \hline
		$M_\text{GUT}$ & $v_{\Sigma_A}$ & $10^{17.8}$ & $\cdot$ & \makecell{$H_A^T$, ${S'_{TT}}^{\textbf{10}}$, ${S'_{DD}}^{\textbf{10}}$, \\ $\Phi^{DT}$, $\Phi^{DD}$, \\ $\Sigma_A^{TT}$, $\Sigma_A^{TD}$, $\Sigma_A^{DT}$} \\ \hline
	\end{tabular}
	\caption{Summary of energy scales, VEV's, heavy fermions, and scalar submultiplets.}
	\label{tab:Escale}
\end{table}
The coefficients $b$ of renormalization group equations are summarized in Table \ref{tab:RCb}.
\begin{table}
	\center
	\begin{tabular}{|c||c|}
		\hline
		Energy scale interval & Coefficients of renormalization group equations \\ \hline
		$M_Z - M_D$ & $b_s = 7$, $b_L = \frac{19}{6}$, $b_Y = -\frac{41}{26}$ \\ \hline
		$M_D - M_R$ & $b_s = \frac{17}{3}$, $b_L = \frac{19}{6}$, $b_Y = -\frac{139}{78}$ \\ \hline
		$M_R - M_S$ & $b_s = \frac{5}{3}$, $b_{2A} = -\frac{1}{2}$, $b_{2B} = \frac{1}{6}$, $b_{B - L} = -\frac{7}{3}$ \\ \hline
		$M_S - M_\Phi$ & $b_s = -\frac{5}{6}$, $b_{2A} = -\frac{1}{2}$, $b_{1A} = -\frac{31}{6}$, $b_{2B} = \frac{1}{6}$, $b_{1B} = -\frac{31}{6}$ \\ \hline
		$M_\Phi - M_\Sigma$ & $b_{3A} = 2$, $b_{2A} = -\frac{7}{2}$, $b_{1A} = -\frac{57}{10}$, $b_{3B} = 1$, $b_{2B} = -\frac{7}{2}$, $b_{1B} = -\frac{57}{10}$ \\ \hline
		$M_\Sigma - M_\text{GUT}$ & $b_{3A} = -\frac{13}{6}$, $b_{2A} = -\frac{15}{2}$, $b_{1A} = -\frac{293}{30}$, $b_{5B} = 4$ \\ \hline
		$M_\text{GUT} -$ & $b_5 = \frac{5}{3}$ \\ \hline
	\end{tabular}
	\caption{Coefficients of renormalization group equations in various energy scale intervals.}
	\label{tab:RCb}
\end{table}
In order to make sure that any current lower bounds on vector-like quark masses are satisfied, we have introduced another energy scale $M_D \sim$ 1 TeV where the masses of two $D$ quarks, the lightest vector-like quarks in this model, appear. With the values of VEV's and masses we have used for unification, the parameters associated with masses of $U$ quarks or neutrinos turn out to have values $|\eta_{H \Phi}| \sim 10^{17}$ GeV, $|\lambda_{H \Sigma \Phi}| \sim 10^{-1}$, and $|\eta_{HS}| \sim |\eta_{SH}| \sim 10^{10}$ GeV. We have assumed $\lambda_U = v_S (= v_\Phi)$ for this estimation.

A few comments are in order on the results of our unification results and their implication: 

\begin{itemize}
\item Clearly, the unification scale is around $10^{18}$ GeV, which means that the proton decay rate contribution from the SU(5)$_A$ sector are much smaller than that of the SU(5)$_B$ sector whose unification scale is around $10^{11}$ GeV. 

\item Since $\eta_{H \Phi}$ is a dimensionful parameter, its value being close to the GUT scale is not unnatural.

\item Within our unification scheme, there are two vector-like down quarks near the TeV scale ($M_D$ in Table \ref{tab:Escale}). These should be accessible at the LHC.

\item The RH gauge boson $W_R$, whose mass is in the $\mathcal{O}(10^2)$ TeV range, is clearly beyond the reach of colliders.

\end{itemize}

\section{Proton lifetime} \label{sec:proton}
In this section, we discuss various proton decay channels and calculate proton lifetime. Since the gauge couplings of SU(5)$_B$ are unified at $M_\Sigma$ which is much lower than $M_\text{GUT}$ as we have noted above, the proton decay processes mediated by the gauge bosons of SU(5)$_B$ are clearly dominant over those by the gauge bosons of SU(5)$_A$. In addition, there are decay channels mediated by $H_B^T$ or $H_A^T$, and the decays by $H_B^T$ are dominant over those by $H_A^T$, since $H_B^T$ is much lighter than $H_A^T$. Hence, we consider only the decay channels mediated by SU(5)$_B$ gauge bosons or $H_B^T$. As mentioned above, for the $H^T_B$-mediated effects to be acceptable, we must put the mass of $H^T_B$ around two orders of magnitude above the SU(5)$_B$ unification scale $M_\Sigma$. For this reason, we have included the threshold corrections due to this scalar multiplet in the gauge coupling evolution (see (\ref{eq:alpha5B})). We find that in any case this effect is very small.

First, we identify the baryon number violating operators due to the gauge boson exchange. Note for this purpose that all the vector-like quarks have the same baryon numbers as the SM quarks, i.e.~$B' = \pm 1/3$. Similarly, the heavy vector-like leptons have the same lepton numbers as the SM leptons, i.e.~$L' = \pm 1$. Prior to spontaneous symmetry breaking, the basic operators that violate the baryon number are $\epsilon_{ijk} \overline{U}_i u^c_j \overline{D}_k e^+$ and $\epsilon_{ijk} \overline{U}_i u^c_j \overline{E^-}d^c_k$ arising from the SU(5)$_B$ gauge boson ${(A_B)^4}_\alpha (= X_{B \alpha})$ exchange, and also $\epsilon_{ijk} \overline{D}_i \nu^c \overline{U}_j d^c_k$ due to the ${(A_B)^5}_\alpha (= Y_{B \alpha})$ exchange.
The corresponding Feynman diagrams are
\begin{center}
	\unitlength = 1 mm
\begin{fmffile}{heavyOFD}
	\qquad \qquad
	\parbox{25 mm}{
		\tiny
		\begin{fmfgraph*}(20,18)
			\fmfstraight
			\fmfset{arrow_len}{2 mm}
			\fmfset{wiggly_len}{2 mm}
			
			\fmfleft{i1,i2}
			\fmfright{o1,o2}
			
			\fmf{fermion,width=0.7}{i1,v1}
			\fmf{fermion,width=0.7}{v1,i2}
			
			\fmf{fermion,width=0.7}{o1,v2}
			\fmf{fermion,width=0.7}{v2,o2}
			
			\fmf{photon,width=0.7,label=$X_B$,label.side=right,label.dist=3}{v1,v2}
			
			\fmfv{label=$U$,label.dist=1}{i1}
			\fmfv{label=$\overline{u^c}$,label.dist=1}{i2}
			\fmfv{label=$\overline{D}$,label.dist=1}{o1}
			\fmfv{label=$e^+$,label.dist=1}{o2}

		\end{fmfgraph*}}
		\qquad \qquad
	\parbox{25 mm}{
		\tiny
		\begin{fmfgraph*}(25,18)
			\fmfstraight
			\fmfset{arrow_len}{2 mm}
			\fmfset{wiggly_len}{2 mm}
			
			\fmfleft{i1,i2}
			\fmfright{o1,o2}
			
			\fmf{fermion,width=0.7}{i1,v1}
			\fmf{fermion,width=0.7}{v1,i2}
			
			\fmf{fermion,width=0.7}{o1,v2}
			\fmf{fermion,width=0.7}{v2,o2}
			
			\fmf{photon,width=0.7,label=$X_B$,label.side=right,label.dist=3}{v1,v2}
			
			\fmfv{label=$U$,label.dist=1}{i1}
			\fmfv{label=$\overline{u^c}$,label.dist=1}{i2}
			\fmfv{label=$\overline{E^-}$,label.dist=1}{o1}
			\fmfv{label=$d^c$,label.dist=1}{o2}
		\end{fmfgraph*}}
		\qquad \qquad
		\parbox{25 mm}{
		\tiny
		\begin{fmfgraph*}(25,18)
			\fmfstraight
			\fmfset{arrow_len}{2 mm}
			\fmfset{wiggly_len}{2 mm}
			
			\fmfleft{i1,i2}
			\fmfright{o1,o2}
			
			\fmf{fermion,width=0.7}{i1,v1}
			\fmf{fermion,width=0.7}{v1,i2}
			
			\fmf{fermion,width=0.7}{o1,v2}
			\fmf{fermion,width=0.7}{v2,o2}
			
			\fmf{photon,width=0.7,label=$Y_B$,label.side=right,label.dist=3}{v1,v2}
			
			\fmfv{label=$U$,label.dist=1}{i1}
			\fmfv{label=$\overline{d^c}$,label.dist=1}{i2}
			\fmfv{label=$\overline{D}$,label.dist=1}{o1}
			\fmfv{label=$\nu^c$,label.dist=1}{o2}
		\end{fmfgraph*}} \newline
\end{fmffile}
\end{center}
In the absence of heavy-light fermion mixing, these operators cannot lead to proton decay. The heavy quark and lepton fields $U, D, E$, however, mix with the light quarks due to the quark and charged lepton seesaw as discussed above, and that in turn leads to proton decay operators involving SM fermion fields. On the other hand, the baryon number violating operators due to the $H_B^T$ exchange have only light SM fermions, and they are directly responsible for proton decay. We present the Feynman diagrams of the SU(5)$_B$ gauge bosons or $H_B^T$ exchange having only light SM external fermions in Sec.~\ref{sec:protoncal}.

\subsection{Mixing parameters $\xi_f$}
In order to explicitly calculate proton lifetime, we assume for simplicity that all the Yukawa coupling matrices are in diagonal forms and $U_f = U_F = \mathbbm{1}$ in (\ref{eq:fFmass}). In this case, the mass eigenstate of each light fermion $f_m$ is a linear combination of single $f_g$ and single $F_g$, and we thus have $f_m \approx f_g + \xi_f F_g$ where $\xi_f$ is the small mixing parameter.
Using (\ref{eq:mixing}), we may write $\xi_f \approx -h_f v_A / (g_f v_C)$ where $h_f$ and $g_f$ are the diagonal elements in the Yukawa coupling matrices $h$ and $g$ associated with the fermion $f$.

Now we express the heavy-light fermion mixing parameters $\xi_f$ in terms of model parameters for a decay process that involves $u, d$ quarks. The generalizations to any other quarks or charged leptons are straightforward. Using (\ref{eq:fFmass}), we can write $u, d$ quark masses as
\begin{align}
	m_u \sim \frac{|h_H'^{11}|^2 |v_{H_A} v_{H_B}|}{|h_S^{11} \lambda_U|}, \qquad
	m_d \sim \frac{|h_H^{11}|^2 |v_{H_A} v_{H_B}|}{|h_\Phi^{11} v_\Phi|}.
	\label{eq:qmass}
\end{align}
The mixing parameters can therefore be expressed in terms of quark masses, VEV's, and Yukawa couplings as
\begin{align}
	|\xi_u| \sim \frac{|h_H'^{11} v_{H_A}|}{|h_S^{11} \lambda_U|} \sim \sqrt{\frac{m_u |v_{H_A}|}{|h_S^{11} v_{H_B} \lambda_U|}}, \qquad
	|\xi_d| \sim \frac{|h_H^{11} v_{H_A}|}{|h_\Phi^{11} v_\Phi|} \sim \sqrt{\frac{m_d |v_{H_A}|}{|h_\Phi^{11} v_{H_B} v_\Phi|}}.
\end{align}
Here, we have eliminated $|h_H'^{11}|$ and $|h_H^{11}|$ to obtain the final expressions of mixing parameters. Since the Wilson coefficient $C^I$ for this decay process has the factor $\xi_u \xi_d$, it is apparent that the larger Yukawa couplings $|h_\Phi^{11} h_S^{11}|$ would produce the larger proton lifetime. However, we have to simultaneously consider all the dominant decay channels as well as all the masses of heavy fermions we have used for unification, and should find the values of Yukawa couplings that satisfy all the conditions and constraints. In addition, we have to make sure that these Yukawa coulings are consistent with the expressions of quark masses given in (\ref{eq:qmass}) because the light fermion masses and the VEV's we assume could require $|h_H'^{11}| > \sqrt{4\pi}$ or $|h_H^{11}| > \sqrt{4\pi}$. In such cases, it is better to eliminate $|h_\Phi^{11}|$ and $|h_S^{11}|$ from the mixing parameters, rather than $|h_H'^{11}|$ and $|h_H^{11}|$ as above, and require large $|h_H'^{11} h_H^{11}|$.

The values of Yukawa couplings we have used are $h_\Phi^{11} = h_\Phi^{22} = M_D / v_\Phi \sim 10^{-6}$, and $h_\Phi^{33} = h_S^{11} = h_S^{22} = h_S^{33} = 1$. We also assumed light fermion masses $m_u = m_d = 10^{-3}$ GeV, $m_s = 10^{-2}$ GeV, $m_e = 10^{-4}$ GeV, and $m_\mu = 10^{-2}$ GeV, which are smaller than the values at the weak scale. We chose these small values to reflect the general tendency of renormalization group running of masses that they decrease as the energy scale increases. We also have used $\lambda_U = v_S (= v_\Phi)$ as before. These masses and Yukawa couplings give $h_H^{11} \sim 10^{-4}$, $h_H^{22} \sim 10^{-4}$, and $h_H^{33} = h_H'^{11} = h_H'^{22} = h_H'^{33} \sim 10^{-1}$, and finally the values of mixing parameters $\xi_u \sim 10^{-8}$, $\xi_d \sim 10^{-5}$, $\xi_s \sim 10^{-5}$, $\xi_e \sim 10^{-9}$, and $\xi_\mu \sim 10^{-8}$.

\subsection{Dominant proton decay channels and predictions on proton lifetime} \label{sec:protoncal}
When a nucleon $N$ decays into a meson $P$ and an anti-lepton $\bar{\ell}$ through an effective interaction induced by dimension-6 operators $\mathcal{O}^I$ $(I = {\Gamma \Gamma'}$ and $\Gamma, \Gamma' = L, R)$, the decay rate is given by \cite{pdmatel}
\begin{align}
	\Gamma \left( N \to \bar{\ell} P \right) = \frac{m_N}{32\pi} \left[ 1 - \left( \frac{m_{P}}{m_N} \right)^2 \right] \left| \sum_I C^I W_0^I \left( N \to P \right) \right|^2.
\end{align}
Here, $C^I$ is the Wilson coefficient of $\mathcal{O}^I$, and $W_0^I \left( N \to P \right)$ is the form factor defined by
\begin{align}
	\langle P (\vec{p})| \mathcal{O}^{\Gamma \Gamma'} |N (\vec{k}, s) \rangle = P_{\Gamma'} \left[ W_0^{\Gamma \Gamma'} (q^2) - \frac{i \slashed{q}}{m_N} W_1^{\Gamma \Gamma'} (q^2) \right] u_N (k, s),
\end{align}
and its value is found from lattice calculation. The proton lifetime is then given by $\tau_p = \hbar / \Gamma (N \to \bar{\ell} P)$.

Now we present the dimension-6 operators, Wilson coefficients, and associated Feynman diagrams for dominant proton decay channels. We assume $m_{X_B} = m_{Y_B}$. \newline

\noindent (i) $p \to e^+ \pi^0, \ \overline{e^-} \pi^0$ \newline
\noindent Here, $e^+$ and $\overline{e^-}$ are the LH and RH positrons, respectively.

\unitlength = 1 mm
\begin{fmffile}{pi0O1FD}
	\parbox{82 mm}{$\mathcal{O}_{\pi^0}^{(1)} = (\overline{u} \gamma^\mu u^c) (\overline{d} \gamma_\mu e^+) \epsilon \epsilon$, \quad
		$C_{\pi^0}^{(1)} = \frac{g_{5B}^2}{m_{X_B}^2} \xi_u \xi_d$,} \quad
	\parbox{20 mm}{
		\tiny
		\begin{fmfgraph*}(20,20)
			\fmfstraight
			\fmfset{arrow_len}{2 mm}
			\fmfset{wiggly_len}{2 mm}
			
			\fmfleft{i1,i2,xi1,xi2,i3,xi3}
			\fmfright{o1,o2,xo1,xo2,o3,xo3}
			
			\fmf{fermion,width=0.7,tension=2}{i2,v1}
			\fmf{fermion,width=0.7,label=$U_1$,label.side=right,label.dist=1.5,tension=2}{v1,v2}
			\fmf{fermion,width=0.7}{v2,i3}
			
			\fmf{fermion,width=0.7,tension=2}{o2,v3}
			\fmf{fermion,width=0.7,label=$\overline{D_1}$,label.side=left,label.dist=1.5,tension=2}{v3,v4}
			\fmf{fermion,width=0.7}{v4,o3}
			
			\fmf{photon,width=0.7,label=$X_B$,label.side=left,label.dist=2}{v2,v4}
			
			\fmf{fermion,width=0.7}{i1,o1}
			
			\fmfv{label=$\xi_u$,label.angle=135,label.dist=1.5,decor.shape=cross,decor.angle=45,decor.size=4}{v1}
			\fmfv{label=$\xi_d$,label.angle=45,label.dist=1.5,decor.shape=cross,decor.angle=45,decor.size=4}{v3}
			
			\fmfv{label=$d$,label.dist=1}{i1}
			\fmfv{label=$d$,label.dist=1}{o1}
			\fmfv{label=$u$,label.dist=1}{i2}
			\fmfv{label=$\overline{u^c}$,label.dist=1}{i3}
			\fmfv{label=$\overline{d}$,label.dist=1}{o2}
			\fmfv{label=$e^+$,label.dist=1}{o3}
		\end{fmfgraph*}}
\end{fmffile}
\newline \newline

\begin{fmffile}{pi0O2FD}
	\parbox{82 mm}{$\mathcal{O}_{\pi^0}^{(2)} = (\overline{u^c} \gamma^\mu u) (\overline{d^c} \gamma_\mu e^-) \epsilon \epsilon$, \quad
		$C_{\pi^0}^{(2)} = \frac{g_{5B}^2}{m_{X_B}^2} \xi_u \xi_e$,} \quad
	\parbox{20 mm}{
		\tiny
		\begin{fmfgraph*}(20,20)
			\fmfstraight
			\fmfset{arrow_len}{2 mm}
			\fmfset{wiggly_len}{2 mm}
			
			\fmfleft{i1,i2,xi1,xi2,i3,xi3}
			\fmfright{o1,o2,xo1,xo2,o3,xo3}
			
			\fmf{fermion,width=0.7,tension=2}{i2,v1}
			\fmf{fermion,width=0.7,label=$U_1$,label.side=right,label.dist=1.5,tension=2}{v1,v2}
			\fmf{fermion,width=0.7}{v2,i3}
			
			\fmf{fermion,width=0.7,tension=2}{o3,v4}
			\fmf{fermion,width=0.7,label=$\overline{E_1^-}$,label.side=right,label.dist=0.5,tension=2}{v4,v3}
			\fmf{fermion,width=0.7}{v3,o2}
			
			\fmf{photon,width=0.7,label=$X_B$,label.side=right,label.dist=2}{v2,v3}
			
			\fmf{fermion,width=0.7}{o1,i1}
			
			\fmfv{label=$\xi_u$,label.angle=135,label.dist=1.5,decor.shape=cross,decor.angle=45,decor.size=4}{v1}
			\fmfv{label=$\xi_e$,label.angle=-45,label.dist=1.5,decor.shape=cross,decor.angle=45,decor.size=4}{v4}
			
			\fmfv{label=$\overline{d^c}$,label.dist=1}{i1}
			\fmfv{label=$\overline{d^c}$,label.dist=1}{o1}
			\fmfv{label=$u$,label.dist=1}{i2}
			\fmfv{label=$\overline{u^c}$,label.dist=1}{i3}
			\fmfv{label=$d^c$,label.dist=1}{o2}
			\fmfv{label=$\overline{e^-}$,label.dist=1}{o3}
		\end{fmfgraph*}}
	\qquad \quad
	\parbox{20 mm}{
		\tiny
		\begin{fmfgraph*}(20,20)
			\fmfstraight
			\fmfset{arrow_len}{2 mm}
			\fmfset{wiggly_len}{2 mm}
			
			\fmfleft{i1,i2,xi1,xi2,i3,xi3}
			\fmfright{o1,o2,xo1,xo2,o3,xo3}
			
			\fmf{fermion,width=0.7,tension=2}{i2,v1}
			\fmf{fermion,width=0.7,label=$U_1$,label.side=right,label.dist=1.5,tension=2}{v1,v2}
			\fmf{fermion,width=0.7}{v2,i3}
			
			\fmf{fermion,width=0.7,tension=2}{o3,v4}
			\fmf{fermion,width=0.7,label=$\overline{E_1^-}$,label.side=right,label.dist=0.5,tension=2}{v4,v3}
			\fmf{fermion,width=0.7}{v3,o2}
			
			\fmf{photon,width=0.7,label=$Y_B$,label.side=right,label.dist=2}{v2,v3}
			
			\fmf{fermion,width=0.7}{o1,i1}
			
			\fmfv{label=$\xi_u$,label.angle=135,label.dist=1.5,decor.shape=cross,decor.angle=45,decor.size=4}{v1}
			\fmfv{label=$\xi_e$,label.angle=-45,label.dist=1.5,decor.shape=cross,decor.angle=45,decor.size=4}{v4}
			
			\fmfv{label=$\overline{u^c}$,label.dist=1}{i1}
			\fmfv{label=$\overline{u^c}$,label.dist=1}{o1}
			\fmfv{label=$u$,label.dist=1}{i2}
			\fmfv{label=$\overline{d^c}$,label.dist=1}{i3}
			\fmfv{label=$u^c$,label.dist=1}{o2}
			\fmfv{label=$\overline{e^-}$,label.dist=1}{o3}
		\end{fmfgraph*}}
\end{fmffile}
\newline \newline

\begin{fmffile}{pi0O3FD}
	\parbox{82 mm}{$\mathcal{O}_{\pi^0}^{(3)} = (u^{c \mathsf{T}} \mathsf{C} d^c) (u^{c \mathsf{T}} \mathsf{C} e^+) \epsilon \epsilon$, \quad
		$C_{\pi^0}^{(3)} = \frac{h_H'^{11*} h_H^{11}}{m_{H_B^T}^2}$,} \quad
	\parbox{20 mm}{
		\tiny
		\begin{fmfgraph*}(20,20)
			\fmfstraight
			\fmfset{arrow_len}{2 mm}
			\fmfset{dash_len}{2 mm}
			
			\fmfleft{i1,i2,xi1,xi2,i3,xi3}
			\fmfright{o1,o2,xo1,xo2,o3,xo3}
			
			\fmf{fermion,width=0.7}{v1,i2}
			\fmf{fermion,width=0.7}{v1,i3}
			
			\fmf{fermion,width=0.7}{v2,o2}
			\fmf{fermion,width=0.7}{v2,o3}
			
			\fmf{dashes,width=0.7,label=$H_B^T$,label.side=right,label.dist=2}{v1,v2}
			
			\fmf{fermion,width=0.7}{o1,i1}
			
			\fmfv{label=$h_H'^{11*}$,label.angle=180,label.dist=2.5}{v1}
			\fmfv{label=$h_H^{11}$,label.angle=0,label.dist=2.5}{v2}
			
			\fmfv{label=$\overline{u^c}$,label.dist=1}{i1}
			\fmfv{label=$\overline{u^c}$,label.dist=1}{o1}
			\fmfv{label=$\overline{u^c}$,label.dist=1}{i2}
			\fmfv{label=$\overline{d^c}$,label.dist=1}{i3}
			\fmfv{label=$u^c$,label.dist=1}{o2}
			\fmfv{label=$e^+$,label.dist=1}{o3}
		\end{fmfgraph*}}
\end{fmffile}
\newline \newline \newline
where $\epsilon$'s are Levi-Civita tensors for SU(3)$_s$ and SU(2)$_w$ indices of the fields in the operators, which we have suppressed for simplicity. 
The decay rate is given by
\begin{align}
	\Gamma \left( p \to e^+ \pi^0, \ \overline{e^-} \pi^0 \right) = \frac{m_p}{32\pi} \left[ 1 - \left( \frac{m_{\pi^0}}{m_p} \right)^2 \right] \left| \left( 2 C_{\pi^0}^{(1)} + 4 C_{\pi^0}^{(2)} + C_{\pi^0}^{(3)} \right) W_0 \left( p \to \pi^0 \right) \right|^2.
\end{align}
Note that we need an additional multiplicative factor 2 for every coefficient $C^I$ corresponding to the operator by the SU(5)$_B$ gauge boson exchange, since $\sigma^\mu_{\alpha \dot{\alpha}} \bar{\sigma}_\mu^{\dot{\beta} \beta} = 2 {\delta_\alpha}^\beta {\delta^{\dot{\beta}}}_{\dot{\alpha}}$. \newline

\noindent (ii) $p \to \mu^+ K^0, \ \overline{\mu^-} K^0$

\begin{fmffile}{K0O1FD}
	\parbox{82 mm}{$\mathcal{O}_{K^0}^{(1)} = (\overline{u} \gamma^\mu u^c) (\overline{s} \gamma_\mu \mu^+) \epsilon \epsilon$, \quad
		$C_{K^0}^{(1)} = \frac{g_{5B}^2}{m_{X_B}^2} \xi_u \xi_s$,} \quad
	\parbox{20 mm}{
		\tiny
		\begin{fmfgraph*}(20,20)
			\fmfstraight
			\fmfset{arrow_len}{2 mm}
			\fmfset{wiggly_len}{2 mm}
			
			\fmfleft{i1,i2,xi1,xi2,i3,xi3}
			\fmfright{o1,o2,xo1,xo2,o3,xo3}
			
			\fmf{fermion,width=0.7,tension=2}{i2,v1}
			\fmf{fermion,width=0.7,label=$U_1$,label.side=right,label.dist=1.5,tension=2}{v1,v2}
			\fmf{fermion,width=0.7}{v2,i3}
			
			\fmf{fermion,width=0.7,tension=2}{o2,v3}
			\fmf{fermion,width=0.7,label=$\overline{D_2}$,label.side=left,label.dist=1.5,tension=2}{v3,v4}
			\fmf{fermion,width=0.7}{v4,o3}
			
			\fmf{photon,width=0.7,label=$X_B$,label.side=left,label.dist=2}{v2,v4}
			
			\fmf{fermion,width=0.7}{i1,o1}
			
			\fmfv{label=$\xi_u$,label.angle=135,label.dist=1.5,decor.shape=cross,decor.angle=45,decor.size=4}{v1}
			\fmfv{label=$\xi_s$,label.angle=45,label.dist=1.5,decor.shape=cross,decor.angle=45,decor.size=4}{v3}
			
			\fmfv{label=$d$,label.dist=1}{i1}
			\fmfv{label=$d$,label.dist=1}{o1}
			\fmfv{label=$u$,label.dist=1}{i2}
			\fmfv{label=$\overline{u^c}$,label.dist=1}{i3}
			\fmfv{label=$\overline{s}$,label.dist=1}{o2}
			\fmfv{label=$\mu^+$,label.dist=1}{o3}
		\end{fmfgraph*}}
\end{fmffile}
\newline \newline

\begin{fmffile}{K0O2FD}
	\parbox{82 mm}{$\mathcal{O}_{K^0}^{(2)} = (\overline{u^c} \gamma^\mu u) (\overline{s^c} \gamma_\mu \mu^-) \epsilon \epsilon$, \quad
		$C_{K^0}^{(2)} = \frac{g_{5B}^2}{m_{X_B}^2} \xi_u \xi_\mu$,} \quad
	\parbox{20 mm}{
		\tiny
		\begin{fmfgraph*}(20,20)
			\fmfstraight
			\fmfset{arrow_len}{2 mm}
			\fmfset{wiggly_len}{2 mm}
			
			\fmfleft{i1,i2,xi1,xi2,i3,xi3}
			\fmfright{o1,o2,xo1,xo2,o3,xo3}
			
			\fmf{fermion,width=0.7,tension=2}{i2,v1}
			\fmf{fermion,width=0.7,label=$U_1$,label.side=right,label.dist=1.5,tension=2}{v1,v2}
			\fmf{fermion,width=0.7}{v2,i3}
			
			\fmf{fermion,width=0.7,tension=2}{o3,v4}
			\fmf{fermion,width=0.7,label=$\overline{E_2^-}$,label.side=right,label.dist=0.5,tension=2}{v4,v3}
			\fmf{fermion,width=0.7}{v3,o2}
			
			\fmf{photon,width=0.7,label=$X_B$,label.side=right,label.dist=2}{v2,v3}
			
			\fmf{fermion,width=0.7}{o1,i1}
			
			\fmfv{label=$\xi_u$,label.angle=135,label.dist=1.5,decor.shape=cross,decor.angle=45,decor.size=4}{v1}
			\fmfv{label=$\xi_\mu$,label.angle=-45,label.dist=1.5,decor.shape=cross,decor.angle=45,decor.size=4}{v4}
			
			\fmfv{label=$\overline{d^c}$,label.dist=1}{i1}
			\fmfv{label=$\overline{d^c}$,label.dist=1}{o1}
			\fmfv{label=$u$,label.dist=1}{i2}
			\fmfv{label=$\overline{u^c}$,label.dist=1}{i3}
			\fmfv{label=$s^c$,label.dist=1}{o2}
			\fmfv{label=$\overline{\mu^-}$,label.dist=1}{o3}
		\end{fmfgraph*}}
\end{fmffile}
\newline \newline
\begin{align}
	\Gamma \left( p \to \mu^+ K^0, \ \overline{\mu^-} K^0 \right) = \frac{m_p}{32\pi} \left[ 1 - \left( \frac{m_{K^0}}{m_p} \right)^2 \right] \left| 2 \left( C_{K^0}^{(1)} + C_{K^0}^{(2)} \right) W_0 \left( p \to K^0 \right) \right|^2. 
\end{align}
\newline

\noindent (iii) $p \to \nu_e^c \pi^+$ \newline

\begin{fmffile}{pi+O1FD}
	\parbox{82 mm}{$\mathcal{O}_{\pi^+}^{(1)} = (\overline{u} \gamma^\mu d^c) (\overline{d} \gamma_\mu \nu_e^c) \epsilon \epsilon$, \quad
		$C_{\pi^+}^{(1)} = \frac{g_{5B}^2}{m_{X_B}^2} \xi_u \xi_d$,} \quad
	\parbox{20 mm}{
		\tiny
		\begin{fmfgraph*}(20,20)
			\fmfstraight
			\fmfset{arrow_len}{2 mm}
			\fmfset{wiggly_len}{2 mm}
			
			\fmfleft{i1,i2,xi1,xi2,i3,xi3}
			\fmfright{o1,o2,xo1,xo2,o3,xo3}
			
			\fmf{fermion,width=0.7,tension=2}{i2,v1}
			\fmf{fermion,width=0.7,label=$U_1$,label.side=right,label.dist=1.5,tension=2}{v1,v2}
			\fmf{fermion,width=0.7}{v2,i3}
			
			\fmf{fermion,width=0.7,tension=2}{o2,v3}
			\fmf{fermion,width=0.7,label=$\overline{D_1}$,label.side=left,label.dist=1.5,tension=2}{v3,v4}
			\fmf{fermion,width=0.7}{v4,o3}
			
			\fmf{photon,width=0.7,label=$Y_B$,label.side=left,label.dist=2}{v2,v4}
			
			\fmf{fermion,width=0.7}{i1,o1}
			
			\fmfv{label=$\xi_u$,label.angle=135,label.dist=1.5,decor.shape=cross,decor.angle=45,decor.size=4}{v1}
			\fmfv{label=$\xi_d$,label.angle=45,label.dist=1.5,decor.shape=cross,decor.angle=45,decor.size=4}{v3}
			
			\fmfv{label=$u$,label.dist=1}{i1}
			\fmfv{label=$u$,label.dist=1}{o1}
			\fmfv{label=$u$,label.dist=1}{i2}
			\fmfv{label=$\overline{d^c}$,label.dist=1}{i3}
			\fmfv{label=$\overline{d}$,label.dist=1}{o2}
			\fmfv{label=$\nu_e^c$,label.dist=1}{o3}
		\end{fmfgraph*}}
\end{fmffile}
\newline \newline

\begin{fmffile}{pi+O2FD}
	\parbox{82 mm}{$\mathcal{O}_{\pi^+}^{(2)} = (u^{c \mathsf{T}} \mathsf{C} d^c) (d^{c \mathsf{T}} \mathsf{C} \nu_e^c) \epsilon \epsilon$, \quad
		$C_{\pi^+}^{(2)} = \frac{h_H'^{11*} h_H^{11}}{m_{H_B^T}^2}$,} \quad
	\parbox{20 mm}{
		\tiny
		\begin{fmfgraph*}(20,20)
			\fmfstraight
			\fmfset{arrow_len}{2 mm}
			\fmfset{dash_len}{2 mm}
			
			\fmfleft{i1,i2,xi1,xi2,i3,xi3}
			\fmfright{o1,o2,xo1,xo2,o3,xo3}
			
			\fmf{fermion,width=0.7}{v1,i2}
			\fmf{fermion,width=0.7}{v1,i3}
			
			\fmf{fermion,width=0.7}{v2,o2}
			\fmf{fermion,width=0.7}{v2,o3}
			
			\fmf{dashes,width=0.7,label=$H_B^T$,label.side=right,label.dist=2}{v1,v2}
			
			\fmf{fermion,width=0.7}{o1,i1}
			
			\fmfv{label=$h_H'^{11*}$,label.angle=180,label.dist=2.5}{v1}
			\fmfv{label=$h_H^{11}$,label.angle=0,label.dist=2.5}{v2}
			
			\fmfv{label=$\overline{u^c}$,label.dist=1}{i1}
			\fmfv{label=$\overline{u^c}$,label.dist=1}{o1}
			\fmfv{label=$\overline{u^c}$,label.dist=1}{i2}
			\fmfv{label=$\overline{d^c}$,label.dist=1}{i3}
			\fmfv{label=$d^c$,label.dist=1}{o2}
			\fmfv{label=$\nu_e^c$,label.dist=1}{o3}
		\end{fmfgraph*}}
\end{fmffile}
\newline \newline
\begin{align}
	\Gamma \left( p \to \nu_e^c \pi^+ \right) = \frac{m_p}{32\pi} \left[ 1 - \left( \frac{m_{\pi^+}}{m_p} \right)^2 \right] \left| \left( 2 C_{\pi^+}^{(1)} + C_{\pi^+}^{(2)} \right) W_0 \left( p \to \pi^+ \right) \right|^2.
\end{align}
\newline

\noindent (iv) $p \to \nu_\mu^c K^+$ \newline

\begin{fmffile}{K+O1FD}
	\parbox{82 mm}{$\mathcal{O}_{K^+}^{(1)} = (\overline{u} \gamma^\mu d^c) (\overline{s} \gamma_\mu \nu_\mu^c) \epsilon \epsilon$, \quad
		$C_{K^+}^{(1)} = \frac{g_{5B}^2}{m_{X_B}^2} \xi_u \xi_s$,} \quad
	\parbox{20 mm}{
		\tiny
		\begin{fmfgraph*}(20,20)
			\fmfstraight
			\fmfset{arrow_len}{2 mm}
			\fmfset{wiggly_len}{2 mm}
			
			\fmfleft{i1,i2,xi1,xi2,i3,xi3}
			\fmfright{o1,o2,xo1,xo2,o3,xo3}
			
			\fmf{fermion,width=0.7,tension=2}{i2,v1}
			\fmf{fermion,width=0.7,label=$U_1$,label.side=right,label.dist=1.5,tension=2}{v1,v2}
			\fmf{fermion,width=0.7}{v2,i3}
			
			\fmf{fermion,width=0.7,tension=2}{o2,v3}
			\fmf{fermion,width=0.7,label=$\overline{D_2}$,label.side=left,label.dist=1.5,tension=2}{v3,v4}
			\fmf{fermion,width=0.7}{v4,o3}
			
			\fmf{photon,width=0.7,label=$Y_B$,label.side=left,label.dist=2}{v2,v4}
			
			\fmf{fermion,width=0.7}{i1,o1}
			
			\fmfv{label=$\xi_u$,label.angle=135,label.dist=1.5,decor.shape=cross,decor.angle=45,decor.size=4}{v1}
			\fmfv{label=$\xi_s$,label.angle=45,label.dist=1.5,decor.shape=cross,decor.angle=45,decor.size=4}{v3}
			
			\fmfv{label=$u$,label.dist=1}{i1}
			\fmfv{label=$u$,label.dist=1}{o1}
			\fmfv{label=$u$,label.dist=1}{i2}
			\fmfv{label=$\overline{d^c}$,label.dist=1}{i3}
			\fmfv{label=$\overline{s}$,label.dist=1}{o2}
			\fmfv{label=$\nu_\mu^c$,label.dist=1}{o3}
		\end{fmfgraph*}}
\end{fmffile}
\newline \newline

\begin{fmffile}{K+O2FD}
	\parbox{82 mm}{$\mathcal{O}_{K^+}^{(2)} = (u^{c \mathsf{T}} \mathsf{C} d^c) (s^{c \mathsf{T}} \mathsf{C} \nu_\mu^c) \epsilon \epsilon$, \quad
		$C_{K^+}^{(2)} = \frac{h_H'^{11*} h_H^{22}}{m_{H_B^T}^2}$,} \quad
	\parbox{20 mm}{
		\tiny
		\begin{fmfgraph*}(20,20)
			\fmfstraight
			\fmfset{arrow_len}{2 mm}
			\fmfset{dash_len}{2 mm}
			
			\fmfleft{i1,i2,xi1,xi2,i3,xi3}
			\fmfright{o1,o2,xo1,xo2,o3,xo3}
			
			\fmf{fermion,width=0.7}{v1,i2}
			\fmf{fermion,width=0.7}{v1,i3}
			
			\fmf{fermion,width=0.7}{v2,o2}
			\fmf{fermion,width=0.7}{v2,o3}
			
			\fmf{dashes,width=0.7,label=$H_B^T$,label.side=right,label.dist=2}{v1,v2}
			
			\fmf{fermion,width=0.7}{o1,i1}
			
			\fmfv{label=$h_H'^{11*}$,label.angle=180,label.dist=2.5}{v1}
			\fmfv{label=$h_H^{11}$,label.angle=0,label.dist=2.5}{v2}
			
			\fmfv{label=$\overline{u^c}$,label.dist=1}{i1}
			\fmfv{label=$\overline{u^c}$,label.dist=1}{o1}
			\fmfv{label=$\overline{u^c}$,label.dist=1}{i2}
			\fmfv{label=$\overline{d^c}$,label.dist=1}{i3}
			\fmfv{label=$s^c$,label.dist=1}{o2}
			\fmfv{label=$\nu_\mu^c$,label.dist=1}{o3}
		\end{fmfgraph*}}
\end{fmffile}
\newline \newline
\begin{align}
	\Gamma \left( p \to \nu_\mu^c K^+ \right) = \frac{m_p}{32\pi} \left[ 1 - \left( \frac{m_{K^+}}{m_p} \right)^2 \right] \left| \left( 2 C_{K^+}^{(1)} + C_{K^+}^{(2)} \right) W_0 \left( p \to K^+ \right) \right|^2. 
\end{align}
\newline

\noindent (v) $p \to e^+ \eta, \ \overline{e^-} \eta$ \newline
\noindent The operators, coefficients, and decay rate are identical to those of $p \to e^+ \pi^0, \ \overline{e^-} \pi^0$ with $\pi^0$ replaced by $\eta$. \newline

\noindent The proton lifetimes from these decay channels are given in Table \ref{tab:taup}. In all the decay modes except for $p \to \mu^+ K^0$, the $H_B^T$-mediated decay processes give the dominant contributions.
\begin{table}
	\center
	\begin{tabular}{|c||c|c|}
		\hline
		Decay channel & Prediction & Current lower bound (90~\% C.L.)~\cite{SuperK} \\ \hline
		$p \to \nu_e^c \pi^+$ & $3.1 \cdot 10^{34}$ yrs. & $3.9 \cdot 10^{32}$ yrs. \\ \hline
		$p \to \nu_\mu^c K^+$ & $3.1 \cdot 10^{34}$ yrs. & $6.6 \cdot 10^{33}$ yrs. \\ \hline
		$p \to e^+ \pi^0, \ \overline{e^-} \pi^0$ & $6.2 \cdot 10^{34}$ yrs. & $1.7 \cdot 10^{34}$ yrs. \\ \hline
		$p \to e^+ \eta, \ \overline{e^-} \eta$ & $4.3 \cdot 10^{36}$ yrs. & $4.2 \cdot 10^{33}$ yrs. \\ \hline
		$p \to \mu^+ K^0, \ \overline{\mu^-} K^0$ & $1.5 \cdot 10^{41}$ yrs. & $6.6 \cdot 10^{33}$ yrs. \\ \hline
	\end{tabular}
	\caption{Proton lifetimes for dominant decay channels in this model. The final states $e^+$ and $\overline{e^-}$ differ only in their helicities. Similarly for muons.}
	\label{tab:taup}
\end{table}

\section{Comments and conclusion} \label{sec:comment}
We have presented a grand unification model based on the gauge group SU(5) $\otimes$ SU(5) $\otimes$ $Z_2$ which predicts three generations of heavy vector-like quarks and leptons. The masses of fields in this model are determined by the requirement of coupling unification and the constraint on proton decay. For instance, the scheme we have presented has two of the down vector-like quarks in the accessible range at the LHC. The proton decay rate appears to put a strong constraint on the unification scheme. After many trials, we succeeded in finding one scheme where the predicted proton lifetimes through various decay channels satisfy the current experimental lower bounds, and we have presented it in this paper. We find that the decay rate of $p \to e^+ \pi^0$ in this model is in the accessible range of planned experiments. For simplicity, we have only chosen the option of having Dirac neutrinos in which case two more decay modes with $\nu^c$ in the final states are possible to give proton lifetimes in the range of $10^{34}$ yrs. Alternatively, one can always extend the model by adding multiplets $(\textbf{15}, \textbf{1}) \oplus (\textbf{1}, \textbf{15})$ which give masses to Majorana neutrinos after spontaneous symmetry breaking. This, however, will affect coupling unification depending on where masses of these new Higgs fields are, and we do not pursue this here. In addition, we note again that our work assumes only one-loop contributions to the $\beta$-functions for coupling unification. Hence, there will be some corrections to our predictions on mass scales and proton lifetime once two-loop effects are included. We do not discuss this here since our goal in this paper has been only to show that a phenomenologically viable unification without SUSY is possible at the leading order. It is also worth pointing out that quark seesaw models with parity provide a simple solution to the strong CP problem without the axion~\cite{BM}. Our model has the potential to embed such a solution into the GUT framework, and this is currently under investigation.  


\section*{Acknowledgement}
This work is supported by the National Science Foundation grant No.~PHY1620074.

\appendix

\section{Scalar potential, vacuum expectation values, and scalar mass spectrum} \label{appendix}
In this appendix, we discuss the scalar sector of the model, and show how the VEV pattern arises. We also present the scalar mass spectrum.

\subsection{Scalar potential}
Since the VEV of $S'$ is assumed to be zero, it does not affect any masses or minimization conditions of the scalar potential.
Its only role is to help achieve grand unification. We therefore neglect $S'$ in the following discussion for simplicity. The scalar potential is
\footnotesize
\begin{align}
	V &= -\mu_H^2 \big[ (H_A^\dagger)_i {H_A}^i + (H_B^\dagger)_\alpha {H_B}^\alpha \big]
	- \mu_\Phi^2 {(\Phi^\dagger)^\alpha}_i {\Phi^i}_\alpha \nonumber \\
	&\quad \quad - \mu_\Sigma^2 \big[ {{\Sigma_A}^i}_j {{\Sigma_A}^j}_i
	+ {{\Sigma_B}^\alpha}_\beta {{\Sigma_B}^\beta}_\alpha \big]
	- \mu_S^2 {(S^\dagger)_{\alpha \beta}}^{ij} {S_{ij}}^{\alpha \beta} \displaybreak[0] \nonumber \\
	&\quad + \eta_{H \Phi} \big[ (H_A^\dagger)_i {H_B}^\alpha {\Phi^i}_\alpha
		+ H_A^i (H_B^\dagger)_\alpha {(\Phi^\dagger)^\alpha}_i \big]
	+ \eta_{H \Sigma} \big[ (H_A^\dagger)_i {H_A}^j {{\Sigma_A}^i}_j
		+ (H_B^\dagger)_\alpha {H_B}^\beta {{\Sigma_B}^\alpha}_\beta \big] \nonumber \\
	&\quad \quad + \eta_\Sigma \big[ {{\Sigma_A}^i}_j {{\Sigma_A}^j}_k {{\Sigma_A}^k}_i
		+ {{\Sigma_B}^\alpha}_\beta {{\Sigma_B}^\beta}_\gamma {{\Sigma_B}^\gamma}_\alpha \big]
	+ \eta_{\Phi \Sigma} \big[ {{\Sigma_A}^i}_j {\Phi^j}_\alpha {(\Phi^\dagger)^\alpha}_i
		+ {{\Sigma_B}^\alpha}_\beta {(\Phi^\dagger)^\beta}_i {\Phi^i}_\alpha \big] \nonumber \\
	&\quad \quad + \eta_{\Phi S} \big[ {\Phi^i}_\alpha {\Phi^j}_\beta {S_{ij}}^{\alpha \beta}
		+ {(\Phi^\dagger)^\alpha}_i {(\Phi^\dagger)^\beta}_j {(S^\dagger)_{\alpha \beta}}^{ij} \big] \nonumber \\
	&\quad \quad + \eta_{S \Phi} \big[ \epsilon^{ijk \ell m} \epsilon_{\alpha \beta \gamma \delta \lambda} {(\Phi^\dagger)^\alpha}_i {S_{jk}}^{\beta \gamma} {S_{\ell m}}^{\delta \lambda}
		+ \epsilon_{ijk \ell m} \epsilon^{\alpha \beta \gamma \delta \lambda} {\Phi^i}_\alpha {(S^\dagger)_{\beta \gamma}}^{jk} {(S^\dagger)_{\delta \lambda}}^{\ell m} \big] \nonumber \\
	&\quad \quad + \eta_{S \Sigma} \big[ {{\Sigma_A}^i}_j {S_{ik}}^{\alpha \beta} {(S^\dagger)_{\alpha \beta}}^{jk}
		+ {{\Sigma_B}^\alpha}_\beta {(S^\dagger)_{\alpha \gamma}}^{ij} {S_{ij}}^{\beta \gamma} \big] \displaybreak[0] \nonumber \\
	&\quad + \lambda_H \big[ (H_A^\dagger)_i {H_A}^i (H_A^\dagger)_j {H_A}^j
		+ (H_B^\dagger)_\alpha {H_B}^\alpha (H_B^\dagger)_\beta {H_B}^\beta \big]
	+ \lambda'_H (H_A^\dagger)_i {H_A}^i (H_B^\dagger)_\alpha {H_B}^\alpha \nonumber \\
	&\quad \quad + \lambda_{H \Phi} \big[ (H_A^\dagger)_i {H_A}^i {(\Phi^\dagger)^\alpha}_j {\Phi^j}_\alpha
		+ (H_B^\dagger)_\alpha {H_B}^\alpha {(\Phi^\dagger)^\beta}_i {\Phi^i}_\beta \big] \nonumber \\
		&\quad \quad \quad + \lambda'_{H \Phi} \big[ (H_A^\dagger)_i {H_A}^j {(\Phi^\dagger)^\alpha}_j {\Phi^i}_\alpha
			+ (H_B^\dagger)_\alpha {H_B}^\beta {(\Phi^\dagger)^\alpha}_i {\Phi^i}_\beta \big] \nonumber \\
	&\quad \quad + \lambda_{H \Sigma} \big[ (H_A^\dagger)_i {H_A}^i {{\Sigma_A}^j}_k {{\Sigma_A}^k}_j
		+ (H_B^\dagger)_\alpha {H_B}^\alpha {{\Sigma_B}^\beta}_\gamma {{\Sigma_B}^\gamma}_\beta \big] \nonumber \\
		&\quad \quad \quad + \lambda'_{H \Sigma} \big[ (H_A^\dagger)_i {H_A}^j {{\Sigma_A}^i}_k {{\Sigma_A}^k}_j
			+ (H_B^\dagger)_\alpha {H_B}^\beta {{\Sigma_B}^\alpha}_\gamma {{\Sigma_B}^\gamma}_\beta \big] \nonumber \\
		&\quad \quad \quad + \lambda''_{H \Sigma} \big[ (H_A^\dagger)_i {H_A}^i {{\Sigma_B}^\alpha}_\beta {{\Sigma_B}^\beta}_\alpha
			+ (H_B^\dagger)_\alpha {H_B}^\alpha {{\Sigma_A}^i}_j {{\Sigma_A}^j}_i \big] \nonumber \displaybreak[0] \\
	&\quad \quad + \lambda_{HS} \big[ (H_A^\dagger)_i {H_A}^i {(S^\dagger)_{\alpha \beta}}^{jk} {S_{jk}}^{\alpha \beta}
		+ (H_B^\dagger)_\alpha {H_B}^\alpha {(S^\dagger)_{\beta \gamma}}^{ij} {S_{ij}}^{\beta \gamma} \big] \nonumber \\
		&\quad \quad \quad + \lambda'_{HS} \big[ (H_A^\dagger)_i {H_A}^j {(S^\dagger)_{\alpha \beta}}^{ik} {S_{jk}}^{\alpha \beta}
			+ (H_B^\dagger)_\alpha {H_B}^\beta {(S^\dagger)_{\beta \gamma}}^{ij} {S_{ij}}^{\alpha \gamma} \big] \nonumber \\
	&\quad \quad + \lambda_{H \Sigma \Phi} \big[ (H_A^\dagger)_i {H_B}^\alpha {{\Sigma_A}^i}_j {\Phi^j}_\alpha
		+ {H_A}^i (H_B^\dagger)_\alpha {{\Sigma_B}^\alpha}_\beta {(\Phi^\dagger)^\beta}_i \big] \nonumber \\
		&\quad \quad \quad + \lambda_{H \Sigma \Phi}^* \big[ {H_A}^i (H_B^\dagger)_\alpha {{\Sigma_A}^j}_i {(\Phi^\dagger)^\alpha}_j
			+ (H_A^\dagger)_i {H_B}^\alpha {{\Sigma_B}^\beta}_\alpha {\Phi^i}_\beta \big] \nonumber \\
	&\quad \quad + \lambda_{H \Phi S} \big[ (H_A^\dagger)_i {H_B}^\alpha {(\Phi^\dagger)^\beta}_j {(S^\dagger)_{\alpha \beta}}^{ij}
		+ {H_A}^i {(H_B^\dagger)}_\alpha {\Phi^j}_\beta {S_{ij}}^{\alpha \beta} \big] \nonumber \displaybreak[0] \\
	&\quad \quad + \lambda_\Phi {(\Phi^\dagger)^\alpha}_i {\Phi^i}_\alpha {(\Phi^\dagger)^\beta}_j {\Phi^j}_\beta
		+ \lambda'_\Phi {(\Phi^\dagger)^\alpha}_i {\Phi^i}_\beta {(\Phi^\dagger)^\beta}_j {\Phi^j}_\alpha \nonumber \\
	&\quad \quad + \lambda_{\Phi \Sigma} \big[ {(\Phi^\dagger)^\alpha}_i {\Phi^i}_\alpha {{\Sigma_A}^j}_k {{\Sigma_A}^k}_j
		+ {(\Phi^\dagger)^\alpha}_i {\Phi^i}_\alpha {{\Sigma_B}^\beta}_\gamma {{\Sigma_B}^\gamma}_\beta \big] \nonumber \\
		&\quad \quad \quad + \lambda'_{\Phi \Sigma} \big[ {(\Phi^\dagger)^\alpha}_i {\Phi^j}_\alpha {{\Sigma_A}^i}_k {{\Sigma_A}^k}_j
			+ {(\Phi^\dagger)^\alpha}_i {\Phi^i}_\beta {{\Sigma_B}^\beta}_\gamma {{\Sigma_B}^\gamma}_\alpha \big]
			+ \lambda''_{\Phi \Sigma} {(\Phi^\dagger)^\alpha}_i {\Phi^j}_\beta {{\Sigma_A}^i}_j {{\Sigma_B}^\beta}_\alpha \nonumber \displaybreak[0] \\
	&\quad \quad + \lambda_{\Phi S} {(\Phi^\dagger)^\alpha}_i {\Phi^i}_\alpha {(S^\dagger)_{\beta \gamma}}^{jk} {S_{jk}}^{\beta \gamma} \nonumber \\
		&\quad \quad \quad + \lambda'_{\Phi S} \big[ {(\Phi^\dagger)^\alpha}_i {\Phi^i}_\beta {(S^\dagger)_{\alpha \gamma}}^{jk} {S_{jk}}^{\beta \gamma}
			+ {(\Phi^\dagger)^\alpha}_i {\Phi^j}_\alpha {(S^\dagger)_{\beta \gamma}}^{ik} {S_{jk}}^{\beta \gamma} \big] \nonumber \\
		&\quad \quad \quad + \lambda''_{\Phi S} {(\Phi^\dagger)^\alpha}_i {\Phi^j}_\beta {(S^\dagger)_{\alpha \gamma}}^{ik} {S_{jk}}^{\beta \gamma} \nonumber \displaybreak[0] \\
%
%
	&\quad \quad + \lambda_{\Phi \Sigma S} \big[ {\Phi^i}_\alpha {\Phi^j}_\beta {{\Sigma_A}^k}_i {S_{jk}}^{\alpha \beta}
		+ {(\Phi^\dagger)^\alpha}_i {(\Phi^\dagger)^\beta}_j {{\Sigma_B}^\gamma}_\alpha {(S^\dagger)_{\beta \gamma}}^{ij} \big] \nonumber \\
		&\quad \quad \quad + \lambda_{\Phi \Sigma S}^* \big[ {(\Phi^\dagger)^\alpha}_i {(\Phi^\dagger)^\beta}_j {{\Sigma_A}^i}_k {(S^\dagger)_{\alpha \beta}}^{jk}
			+ {\Phi^i}_\alpha {\Phi^j}_\beta {{\Sigma_B}^\alpha}_\gamma {S_{ij}}^{\beta \gamma} \big] \nonumber \\
	&\quad \quad + \lambda_{S \Phi \Sigma} \big[ \epsilon^{jk \ell mn} \epsilon_{\alpha \beta \gamma \delta \lambda} {(\Phi^\dagger)^\alpha}_i {{\Sigma_A}^i}_j {S_{k \ell}}^{\beta \gamma} {S_{mn}}^{\delta \lambda}
		+ \epsilon_{ijk \ell m} \epsilon^{\beta \gamma \delta \lambda \mu} {\Phi^i}_\alpha {{\Sigma_B}^\alpha}_\beta {(S^\dagger)_{\gamma \delta}}^{jk}{(S^\dagger)_{\lambda \mu}}^{\ell m} \big] \nonumber \\
		&\quad \quad \quad + \lambda_{S \Phi \Sigma}^* \big[ \epsilon_{jk \ell mn} \epsilon^{\alpha \beta \gamma \delta \lambda} {\Phi^i}_\alpha {{\Sigma_A}^j}_i {(S^\dagger)_{\beta \gamma}}^{k \ell}{(S^\dagger)_{\delta \lambda}}^{mn}
			+ \epsilon^{ijk \ell m} \epsilon_{\beta \gamma \delta \lambda \mu} {(\Phi^\dagger)^\alpha}_i {{\Sigma_B}^\beta}_\alpha {S_{jk}}^{\gamma \delta} {S_{\ell m}}^{\lambda \mu} \big] \nonumber \\
		&\quad \quad \quad + \lambda'_{S \Phi \Sigma} \big[ \epsilon^{ik \ell mn} \epsilon_{\alpha \beta \gamma \delta \lambda} {(\Phi^\dagger)^\alpha}_i {{\Sigma_A}^j}_k {S_{j \ell}}^{\beta \gamma} {S_{mn}}^{\delta \lambda}
			+ \epsilon_{ijk \ell m} \epsilon^{\alpha \gamma \delta \lambda \mu} {\Phi^i}_\alpha {{\Sigma_B}^\beta}_\gamma {(S^\dagger)_{\beta \delta}}^{jk} {(S^\dagger)_{\lambda \mu}}^{\ell m} \big] \nonumber \\
		&\quad \quad \quad + \lambda_{S \Phi \Sigma}'^* \big[ \epsilon_{ik \ell mn} \epsilon^{\alpha \beta \gamma \delta \lambda} {\Phi^i}_\alpha {{\Sigma_A}^k}_j {(S^\dagger)_{\beta \gamma}}^{j \ell} {(S^\dagger)_{\delta \lambda}}^{mn}
			+ \epsilon^{ijk \ell m} \epsilon_{\alpha \gamma \delta \lambda \mu} {(\Phi^\dagger)^\alpha}_i {{\Sigma_B}^\gamma}_\beta {S_{jk}}^{\beta \delta} {S_{\ell m}}^{\lambda \mu} \big] \nonumber \\
	&\quad \quad + \lambda_{S \Phi} \big[ \epsilon^{jk \ell mn} \epsilon_{\beta \gamma \delta \lambda \mu} {\Phi^i}_\alpha {S_{ij}}^{\alpha \beta} {S_{k \ell}}^{\gamma \delta} {S_{mn}}^{\lambda \mu}
		+ \epsilon_{jk \ell mn} \epsilon^{\beta \gamma \delta \lambda \mu} {(\Phi^\dagger)^\alpha}_i {(S^\dagger)_{\alpha \beta}}^{ij} {(S^\dagger)_{\gamma \delta}}^{k \ell} {(S^\dagger)_{\lambda \mu}}^{mn} \big] \nonumber \\
		&\quad \quad \quad + \lambda'_{S \Phi} \big[ \epsilon^{jk \ell mn} \epsilon_{\beta \gamma \delta \lambda \mu} {\Phi^i}_\alpha {S_{ij}}^{\beta \gamma} {S_{k \ell}}^{\alpha \delta} {S_{mn}}^{\lambda \mu}
		+ \epsilon_{jk \ell mn} \epsilon^{\beta \gamma \delta \lambda \mu} {(\Phi^\dagger)^\alpha}_i {(S^\dagger)_{\beta \gamma}}^{ij} {(S^\dagger)_{\alpha \delta}}^{k \ell} {(S^\dagger)_{\lambda \mu}}^{mn} \big] \nonumber \displaybreak[0] \\
	&\quad \quad + \lambda_{\Sigma} \big[ {{\Sigma_A}^i}_j {{\Sigma_A}^j}_i {{\Sigma_A}^k}_\ell {{\Sigma_A}^\ell}_k
		+ {{\Sigma_B}^\alpha}_\beta {{\Sigma_B}^\beta}_\alpha {{\Sigma_B}^\gamma}_\delta {{\Sigma_B}^\delta}_\gamma \big] \nonumber \\
		&\quad \quad \quad + \lambda'_{\Sigma} \big[ {{\Sigma_A}^i}_j {{\Sigma_A}^j}_k {{\Sigma_A}^k}_\ell {{\Sigma_A}^\ell}_i
			+ {{\Sigma_B}^\alpha}_\beta {{\Sigma_B}^\beta}_\gamma {{\Sigma_B}^\gamma}_\delta {{\Sigma_B}^\delta}_\alpha \big]
		+ \lambda''_{\Sigma} {{\Sigma_A}^i}_j {{\Sigma_A}^j}_i {{\Sigma_B}^\alpha}_\beta {{\Sigma_B}^\beta}_\alpha \nonumber \displaybreak[0] \\
	&\quad \quad + \lambda_{\Sigma S} \big[ {{\Sigma_A}^i}_j {{\Sigma_A}^j}_i {(S^\dagger)_{\alpha \beta}}^{k \ell} {S_{k \ell}}^{\alpha \beta}
		+ {{\Sigma_B}^\alpha}_\beta {{\Sigma_B}^\beta}_\alpha {(S^\dagger)_{\gamma \delta}}^{ij} {S_{ij}}^{\gamma \delta} \big] \nonumber \\
		&\quad \quad \quad + \lambda'_{\Sigma S} \big[ {{\Sigma_A}^i}_j {{\Sigma_A}^j}_k {(S^\dagger)_{\alpha \beta}}^{k \ell} {S_{i \ell}}^{\alpha \beta}
			+ {{\Sigma_B}^\alpha}_\beta {{\Sigma_B}^\beta}_\gamma {(S^\dagger)_{\alpha \delta}}^{ij} {S_{ij}}^{\gamma \delta} \big] \nonumber \\
		&\quad \quad \quad + \lambda''_{\Sigma S} \big[ {{\Sigma_A}^i}_j {{\Sigma_A}^k}_\ell {(S^\dagger)_{\alpha \beta}}^{j \ell} {S_{ik}}^{\alpha \beta}
			+ {{\Sigma_B}^\alpha}_\beta {{\Sigma_B}^\gamma}_\delta {(S^\dagger)_{\alpha \gamma}}^{ij} {S_{ij}}^{\beta \delta} \big] \nonumber \\
		&\quad \quad \quad + \lambda'''_{\Sigma S} {{\Sigma_A}^i}_j {{\Sigma_B}^\alpha}_\beta {(S^\dagger)_{\alpha \delta}}^{jk} {S_{ki}}^{\delta \beta} \nonumber \displaybreak[0] \\
	&\quad \quad + \lambda_{S} {(S^\dagger)_{\alpha \beta}}^{ij} {S_{ij}}^{\alpha \beta} {(S^\dagger)_{\gamma \delta}}^{k \ell} {S_{k \ell}}^{\gamma \delta} \nonumber \\
		&\quad \quad \quad + \lambda'_{S} \big[ {(S^\dagger)_{\alpha \beta}}^{ij} {S_{jk}}^{\alpha \beta} {(S^\dagger)_{\gamma \delta}}^{k \ell} {S_{\ell i}}^{\gamma \delta}	
			+ {(S^\dagger)_{\alpha \beta}}^{ij} {S_{ij}}^{\beta \gamma} {(S^\dagger)_{\gamma \delta}}^{k \ell} {S_{k \ell}}^{\delta \alpha}
 \big] \nonumber \\
		&\quad \quad \quad + \lambda''_{S} {(S^\dagger)_{\alpha \beta}}^{ij} {S_{k \ell}}^{\alpha \beta} {(S^\dagger)_{\gamma \delta}}^{k \ell} {S_{ij}}^{\gamma \delta}
			+ \lambda'''_{S} {(S^\dagger)_{\alpha \beta}}^{ij} {S_{jk}}^{\beta \gamma} {(S^\dagger)_{\gamma \delta}}^{k \ell} {S_{\ell i}}^{\delta \alpha}.
\end{align}
\normalsize
Note that all the coupling constants except for $\lambda_{H \Sigma \Phi}$, $\lambda_{\Phi \Sigma S}$, $\lambda_{S \Phi \Sigma}$, and $\lambda'_{S \Phi \Sigma}$ are real due to the $Z_2$ symmetry and Hermiticity of $V$.

\subsection{Vacuum expectation values}
The VEV's that satisfy the condition $v_{\Sigma_A} \gg v_{\Sigma_B} \gg |v_\Phi| \sim |v_S| \gg |v_{H_B}| \gg |v_{H_A}|$ are approximately written as
\small
\begin{align}
	v_{\Sigma_A}^2 &\approx \frac{\mu_\Sigma^2}{2 (30 \lambda_\Sigma + 7 \lambda'_\Sigma)}, \displaybreak[0] \\
	v_{\Sigma_B}^2 &\approx \frac{30 \lambda_\Sigma + 7 \lambda'_\Sigma - 30 \lambda''_\Sigma}{2 (30 \lambda_\Sigma + 7 \lambda'_\Sigma)} v_{\Sigma_A}^2, \displaybreak[0] \\
	|v_\Phi|^2 &\approx \frac{\mu_\Phi^2 - 2 \eta_{\Phi \Sigma} \big( v_{\Sigma_A} + v_{\Sigma_B} \big) - 2 (15 \lambda_{\Phi \Sigma} + 2 \lambda'_{\Phi \Sigma}) \big( v_{\Sigma_A}^2 + v_{\Sigma_B}^2 \big) - 4 \lambda''_{\Phi \Sigma} v_{\Sigma_A} v_{\Sigma_B}}{2 (3 \lambda_\Phi + \lambda'_\Phi)}, \displaybreak[0] \\
	|v_S|^2 &\approx \frac{\mu_S^2 + 3 \eta_{S \Sigma} \big( v_{\Sigma_A} + v_{\Sigma_B} \big) - 3 (10 \lambda_{\Sigma S} + 3 \lambda'_{\Sigma S} + 3 \lambda''_{\Sigma S}) \big( v_{\Sigma_A}^2 + v_{\Sigma_B}^2 \big) - 9 \lambda'''_{\Sigma S} v_{\Sigma_A} v_{\Sigma_B}}{4 (4 \lambda_S + 4 \lambda'_S + 4 \lambda''_S + \lambda'''_S)}, \displaybreak[0] \\
	|v_{H_B}|^2 &\approx \frac{1}{2 \lambda_H} [\mu_H^2 + 3 \eta_{H \Sigma} v_{\Sigma_B} - 3 (10 \lambda_{H \Sigma} + 3 \lambda'_{H \Sigma}) v_{\Sigma_B}^2 - 30 \lambda''_{H \Sigma} v_{\Sigma_A}^2 \nonumber \\
		&\quad \quad \quad - 3 \lambda_{H \Phi} |v_\Phi|^2 - 2 (2 \lambda_{HS} + \lambda'_{HS}) |v_S|^2], \displaybreak[0] \\
	|v_{H_A}|^2 &\approx \frac{1}{2 \lambda_H} [\mu_H^2 + 3 \eta_{H \Sigma} v_{\Sigma_A} - 3 (10 \lambda_{H \Sigma} + 3 \lambda'_{H \Sigma}) v_{\Sigma_A}^2 - 30 \lambda''_{H \Sigma} v_{\Sigma_B}^2 \nonumber \\
		&\quad \quad \quad  - 3 \lambda_{H \Phi} |v_\Phi|^2 - 2 (2 \lambda_{HS} + \lambda'_{HS}) |v_S|^2 - \lambda'_H |v_{H_B}|^2].
\end{align}
\normalsize
The minimization conditions of the scalar potential are
\small
\begin{align}
	&-2 \mu_\Sigma^2 - 3 \eta_\Sigma v_{\Sigma_A} + 4 (30 \lambda_\Sigma + 7 \lambda'_\Sigma) v_{\Sigma_A}^2 \approx 0, \displaybreak[0] \\
	&-2 \mu_\Sigma^2 - 3 \eta_\Sigma v_{\Sigma_B} + 4 (30 \lambda_\Sigma + 7 \lambda'_\Sigma) v_{\Sigma_B}^2 + 60 \lambda''_\Sigma v_{\Sigma_A}^2 \approx 0, \displaybreak[0] \\
	&-\mu_\Phi^2 + 2 \eta_{\Phi \Sigma} \big( v_{\Sigma_A} + v_{\Sigma_B} \big) + 2 (3 \lambda_\Phi + \lambda'_\Phi) |v_\Phi|^2 \nonumber \\
		&\quad \quad + 2 (15 \lambda_{\Phi \Sigma} + 2 \lambda'_{\Phi \Sigma}) \big( v_{\Sigma_A}^2 + v_{\Sigma_B}^2 \big) + 4 \lambda''_{\Phi \Sigma} v_{\Sigma_A} v_{\Sigma_B} \approx 0,\displaybreak[0]  \\
	&-\mu_S^2 - 3 \eta_{S \Sigma} \big( v_{\Sigma_A} + v_{\Sigma_B} \big) + 3 (10 \lambda_{\Sigma S} + 3 \lambda'_{\Sigma S} + 3 \lambda''_{\Sigma S}) \big( v_{\Sigma_A}^2 + v_{\Sigma_B}^2 \big) \nonumber \\
		&\quad \quad + 9 \lambda'''_{\Sigma S} v_{\Sigma_A} v_{\Sigma_B} + 4 (4 \lambda_S + 4 \lambda'_S + 4 \lambda''_S + \lambda'''_S) |v_S|^2 \approx 0, \displaybreak[0] \\
	&-\mu_H^2 - 3 \eta_{H \Sigma} v_{\Sigma_B} + 2 \lambda_H |v_{H_B}|^2 + 3 \lambda_{H \Phi} |v_\Phi|^2 \nonumber \\
		&\quad \quad + 3 (10 \lambda_{H \Sigma} + 3 \lambda'_{H \Sigma}) v_{\Sigma_B}^2 + 30 \lambda''_{H \Sigma} v_{\Sigma_A}^2 + 2 (2 \lambda_{HS} + \lambda'_{HS}) |v_S|^2 \approx 0, \displaybreak[0] \\
	&-\mu_H^2 - 3 \eta_{H \Sigma} v_{\Sigma_A} + 2 \lambda_H |v_{H_A}|^2 + \lambda'_H |v_{H_B}|^2 + 3 \lambda_{H \Phi} |v_\Phi|^2 \nonumber \\
	&\quad \quad + 3 (10 \lambda_{H \Sigma} + 3 \lambda'_{H \Sigma}) v_{\Sigma_A}^2 + 30 \lambda''_{H \Sigma} v_{\Sigma_B}^2 + 2 (2 \lambda_{HS} + \lambda'_{HS}) |v_S|^2 \approx 0,
\end{align}
\normalsize
and
\small
\begin{gather}
	v_{\Sigma_A}, v_{\Sigma_B} > 0, \quad
	v_{\Sigma_A}, v_{\Sigma_B} > \frac{3 \eta_\Sigma}{8 (30 \lambda_\Sigma + 7 \lambda'_\Sigma)}, \quad
	3 \lambda_\Phi + \lambda'_\Phi > 0, \nonumber \\
	4 \lambda_S + 4 \lambda'_S + 4 \lambda''_S + \lambda'''_S > 0, \quad
	\lambda_H > 0
\end{gather}
\normalsize
when $30 \lambda_\Sigma + 7 \lambda'_\Sigma > 0$.

\subsection{Scalar mass spectrum}
The masses of physical scalar fields are
\small
\begin{align}
	m_{\Sigma_{A1}}^2 &\approx 240 (30 \lambda_\Sigma + 7 \lambda'_\Sigma) v_{\Sigma_A}^2
		\approx 120 \mu_\Sigma^2, \\
	m_{\Sigma_{B1}}^2 &\approx 240 (30 \lambda_\Sigma + 7 \lambda'_\Sigma) v_{\Sigma_B}^2, \\
	m_{\Phi^T_1}^2 &= 6 (3 \lambda_\Phi + \lambda'_\Phi) |v_\Phi|^2, \\
	m_{{S_1}^1}^2 &= 8 (4 \lambda_S + 4 \lambda'_S + 4 \lambda''_S + \lambda'''_S) |v_S|^2, \\
	m_{\phi_{H_B}}^2 &= 2 \lambda_H |v_{H_B}|^2, \\
	m_{\phi_{H_A}}^2 &= 2 \lambda_H |v_{H_A}|^2
\end{align}
\normalsize
and also roughly
\small
\begin{align}
	m_{\phi_{\Sigma_{A8}^T}}^2 &\sim -\mu_\Sigma^2 + 4 (19 \lambda_\Sigma + 6 \lambda_\Sigma') v_{\Sigma_A}^2, \displaybreak[0] \\
	m_{\phi_{\Sigma_{A3}^D}}^2 &\sim -\mu_\Sigma^2 - 6 \eta_\Sigma v_{\Sigma_A} + \lambda_{H \Sigma} |v_{H_B}|^2 + 3 \lambda_{\Phi \Sigma} |v_\Phi|^2 \nonumber \\
		&\quad \quad + 6 (16 \lambda_\Sigma + 9 \lambda_\Sigma') v_{\Sigma_A}^2 + 30 \lambda_\Sigma'' v_{\Sigma_B}^2 + 2 (2 \lambda_{\Sigma S} + \lambda_{\Sigma S}') |v_S|^2, \displaybreak[0] \\
	m_{\phi_{\Sigma_{B8}^T}}^2 &\sim -\mu_\Sigma^2 - 6 \eta_\Sigma v_{\Sigma_B} + (3 \lambda_{\Phi \Sigma} + \lambda_{\Phi \Sigma}') |v_\Phi|^2 \nonumber \\
		&\quad \quad + 4 (19 \lambda_\Sigma + 6 \lambda_\Sigma') v_{\Sigma_B}^2 + 30 \lambda_\Sigma'' v_{\Sigma_A}^2 + 4 \lambda_{\Phi S} |v_S|^2, \displaybreak[0] \\
	m_{\phi_{\Sigma_{B3}^D}}^2 &\sim -\mu_\Sigma^2 - 6 \eta_\Sigma v_{\Sigma_B} + 3 \lambda_{\Phi \Sigma} |v_\Phi|^2 \nonumber \\
		&\quad \quad + 6 (16 \lambda_\Sigma + 9 \lambda_\Sigma') v_{\Sigma_B}^2 + 30 \lambda_\Sigma'' v_{\Sigma_A}^2 + 2 (2 \lambda_{\Sigma S} + \lambda_{\Sigma S}') |v_S|^2, \displaybreak[0] \\
	m_{\phi_{\Phi^T_8}}^2 &\sim -\mu_\Phi^2 + 2 \eta_{\Phi \Sigma} (v_{\Sigma_A} + v_{\Sigma_B}) + 2 (3 \lambda_\Phi + 2 \lambda_\Phi' - 6 \lambda_{\Phi S} + \lambda_{\Phi S}') |v_\Phi|^2 \nonumber \\
		&\quad \quad + 2 (15 \lambda_{\Phi \Sigma} + 2 \lambda_{\Phi \Sigma}') (v_{\Sigma_A}^2 + v_{\Sigma_B}^2) + 4 \lambda_{\Phi \Sigma}'' v_{\Sigma_A} v_{\Sigma_B} - 4 \lambda_{\Phi S} |v_S|^2, \displaybreak[0] \\
	m_{\phi_{\Phi^{TD}}}^2 &\sim -\mu_\Phi^2 + \eta_{\Phi \Sigma} (2 v_{\Sigma_A} - 3 v_{\Sigma_B}) + 2 (15 \lambda_{\Phi \Sigma} + 2 \lambda_{\Phi \Sigma}') v_{\Sigma_A}^2 \nonumber \\
		&\quad \quad + 3 (10 \lambda_{\Phi \Sigma} + 3 \lambda_{\Phi \Sigma}') v_{\Sigma_B}^2 - 6 \lambda_{\Phi \Sigma}'' v_{\Sigma_A} v_{\Sigma_B}, \displaybreak[0] \\
	m_{\phi_{\Phi^{DT}}}^2 &\sim -\mu_\Phi^2 - 3 \eta_{\Phi \Sigma} v_{\Sigma_A} + 3 (10 \lambda_{\Phi \Sigma} + 3 \lambda_{\Phi \Sigma}') v_{\Sigma_A}^2, \displaybreak[0] \\
	m_{\phi_{\Phi^{DD}}}^2 &\sim -\mu_\Phi^2 - 3 \eta_{\Phi \Sigma} v_{\Sigma_A} + 3 (10 \lambda_{\Phi \Sigma} + 3 \lambda_{\Phi \Sigma}') v_{\Sigma_A}^2, \displaybreak[0] \\
	m_{\phi_{{S_{\bar{3}}}^3}}^2 &\sim -4 \mu_S^2 + 8 \eta_{S \Sigma} (v_{\Sigma_A} + v_{\Sigma_B}) + 2 (-6 \lambda_{\Phi S} + 4 \lambda_{\Phi S}' + \lambda_{\Phi S}'') |v_\Phi|^2 \nonumber \\
		&\quad \quad + 8 (15 \lambda_{\Sigma S} + 2 \lambda_{\Sigma S}') (v_{\Sigma_A}^2 + v_{\Sigma_B}^2) + 16 \lambda_{\Sigma S}''' v_{\Sigma_A} v_{\Sigma_B} + 32 \lambda_S |v_S|^2, \displaybreak[0] \\
	m_{\phi_{{S_{TT}}^{TD}}}^2 &\sim -4 \mu_S^2 + 2 \eta_{S \Sigma} (4 v_{\Sigma_A} - v_{\Sigma_B}) + 8 (15 \lambda_{\Sigma S} + 2 \lambda_{\Sigma S}' + 2 \lambda_{\Sigma S}'') v_{\Sigma_A}^2 \nonumber \\
		&\quad \quad + 2 (60 \lambda_{\Sigma S} + 13 \lambda_{\Sigma S}' - 12 \lambda_{\Sigma S}'') v_{\Sigma_B}^2 - 4 \lambda_{\Sigma S}''' v_{\Sigma_A} v_{\Sigma_B}, \displaybreak[0] \\
	m_{\phi_{{S_{TT}}^{DD}}}^2 &\sim -4 \mu_S^2 + 4 \eta_{S \Sigma} (2 v_{\Sigma_A} - 3 v_{\Sigma_B}) + 8 (15 \lambda_{\Sigma S} + 2 \lambda_{\Sigma S}' + 2 \lambda_{\Sigma S}'') v_{\Sigma_A}^2 \nonumber \\
		&\quad \quad + 12 (10 \lambda_{\Sigma S} + 3 \lambda_{\Sigma S}' + 3 \lambda_{\Sigma S}'') v_{\Sigma_B}^2 - 24 \lambda_{\Sigma S}''' v_{\Sigma_A} v_{\Sigma_B} \nonumber \\
		&\quad \quad + 16 (2 \lambda_S + \lambda_S' + 2 \lambda_S'') |v_S|^2, \displaybreak[0] \\
	m_{\phi_{{S_{TD}}^{TT}}}^2 &\sim -4 \mu_S^2 + 2 \eta_{S \Sigma} (4 v_{\Sigma_B} - v_{\Sigma_A}) + 2 (60 \lambda_{\Sigma S} + 13 \lambda_{\Sigma S}' - 12 \lambda_{\Sigma S}'') v_{\Sigma_A}^2 \nonumber \\
		&\quad \quad + 8 (15 \lambda_{\Sigma S} + 2 \lambda_{\Sigma S}' + 2 \lambda_{\Sigma S}'') v_{\Sigma_B}^2 - 4 \lambda_{\Sigma S}''' v_{\Sigma_A} v_{\Sigma_B}, \displaybreak[0] \\
	m_{\phi_{{S_{TD}}^{TD}}}^2 &\sim -4 \mu_S^2 - 2 \eta_{S \Sigma} (v_{\Sigma_A} + v_{\Sigma_B}) + 4 \lambda_{HS} |v_{H_B}|^2 + (-12 \lambda_{\Phi S} + 4 \lambda_{\Phi S}' + \lambda_{\Phi S}'') |v_\Phi|^2 \nonumber \\
		&\quad \quad + 2 (60 \lambda_{\Sigma S} + 13 \lambda_{\Sigma S}' - 12 \lambda_{\Sigma S}'') (v_{\Sigma_A}^2 + v_{\Sigma_B}^2) + \lambda_{\Sigma S}''' v_{\Sigma_A} v_{\Sigma_B} \nonumber \\
		&\quad \quad + 4 (8 \lambda_S + 4 \lambda_S' + \lambda_S'') |v_S|^2, \displaybreak[0] \\
	m_{\phi_{{S_{TD}}^{DD}}}^2 &\sim -4 \mu_S^2 - 2 \eta_{S \Sigma} (v_{\Sigma_A} + 6 v_{\Sigma_B}) + 2 (60 \lambda_{\Sigma S} + 13 \lambda_{\Sigma S}' - 12 \lambda_{\Sigma S}'') v_{\Sigma_A}^2 \nonumber \\
		&\quad \quad + 12 (10 \lambda_{\Sigma S} + 3 \lambda_{\Sigma S}' + 3 \lambda_{\Sigma S}'') v_{\Sigma_B}^2 + 6 \lambda_{\Sigma S}''' v_{\Sigma_A} v_{\Sigma_B}, \displaybreak[0] \\
	m_{\phi_{{S_{DD}}^{TT}}}^2 &\sim -4 \mu_S^2 + 4 \eta_{S \Sigma} (2 v_{\Sigma_B} - 3 v_{\Sigma_A}) + 12 (10 \lambda_{\Sigma S} + 3 \lambda_{\Sigma S}' + 3 \lambda_{\Sigma S}'') v_{\Sigma_A}^2 \nonumber \\
		&\quad \quad + 8 (15 \lambda_{\Sigma S} + 2 \lambda_{\Sigma S}' + 2 \lambda_{\Sigma S}'') v_{\Sigma_B}^2 - 24 \lambda_{\Sigma S}''' v_{\Sigma_A} v_{\Sigma_B}, \displaybreak[0] \\
	m_{\phi_{{S_{DD}}^{TD}}}^2 &\sim -4 \mu_S^2 - 2 \eta_{S \Sigma} (6 v_{\Sigma_A} + v_{\Sigma_B}) + 12 (10 \lambda_{\Sigma S} + 3 \lambda_{\Sigma S}' + 3 \lambda_{\Sigma S}'') v_{\Sigma_A}^2 \nonumber \\
		&\quad \quad + 2 (60 \lambda_{\Sigma S} + 13 \lambda_{\Sigma S}' - 12 \lambda_{\Sigma S}'') v_{\Sigma_B}^2 + 6 \lambda_{\Sigma S}''' v_{\Sigma_A} v_{\Sigma_B}, \displaybreak[0] \\
	m_{\phi_{H_B^T}}^2 &\sim -\mu_H^2 + 2 \eta_{H \Sigma} v_{\Sigma_B} + 2 (15 \lambda_{H \Sigma} + 2 \lambda_{H \Sigma}') v_{\Sigma_B}^2 + 20 \lambda_{H \Sigma}'' v_{\Sigma_A}^2, \displaybreak[0] \\
	m_{\phi_{H_A^T}}^2 &\sim -\mu_H^2 + 2 \eta_{H \Sigma} v_{\Sigma_A} + 2 (15 \lambda_{H \Sigma} + 2 \lambda_{H \Sigma}') v_{\Sigma_A}^2.
\end{align}
\normalsize
We obtained these rough expressions by setting all the fields to their VEV's except for the single field in the multiplet of interest. As a result, any mixing terms among fields are ignored even when their coefficients are the same as those of non-mixing terms we have kept. We also do not present the masses of fields in $S'$.

Note that it is possible to have completely different parametrizations of VEV's and masses as well as different minimization conditions. For example, very large values of $\eta_\Sigma$ are incompatible with the minimization conditions on $v_{\Sigma_A}$ and $v_{\Sigma_B}$ presented above if $30 \lambda_\Sigma + 7 \lambda'_\Sigma > 0$. However, they are allowed if $30 \lambda_\Sigma + 7 \lambda'_\Sigma < 0$.




\end{document}